\documentclass[twocolumn,showpacs,preprintnumbers,amsmath,amssymb]{revtex4}
\usepackage{graphicx}% Include figure files
\usepackage{dcolumn}% Align table columns on decimal point
\usepackage{bm}% bold math
\usepackage{hyperref}

\begin{document}

%\preprint{APS/123-QED}

%\title{Derivation of the Bohr ground-state atomic radius from spin-orbit interaction and hypothesis of angular momentum %constancy}% Force line breaks with \\

%\title{Spin-orbit interaction dynamics of atomic hydrogen in classical electrodynamics}% Force line breaks %with \\

\title{Regarding Llewellyn Thomas's paper of 1927 and the ``hidden momentum'' of a magnetic dipole in an electric field}% Force line breaks with \\

\author{David C. Lush } 
% \altaffiliation[Also at ]{Physics Department, XYZ University.}%Lines break automatically or can be forced with \\
%\author{Second Author}%
% \email{Second.Author@institution.edu}
\affiliation{%
%Authors' institution and/or address\\
%This line break forced with \textbackslash\textbackslash
 d.lush@comcast.net \\}%

%\homepage{http://www.Second.institution.edu/~Charlie.Author}

\date{\today}% It is always \today, today,
             %  but any date may be explicitly specified

\begin{abstract} 

L. H. Thomas, in his 1927 paper, ``The Kinematics of an Electron with an Axis'', explained the then-anomalous factor of one-half in atomic spin-orbit coupling as due to a relativistic precession of the electron spin axis.  Thomas's explanation required also that the total of the orbit-averaged, or ``secular'', orbital and spin angular momenta of the electron be a conserved quantity, as he found to be the case for either of two possible equations of translational motion of the magnetic electron.  Thomas's finding is seen in the present work to require the ``hidden momentum'' of the electron intrinsic magnetic moment in the Coulomb field of the proton be omitted from its equation of translational motion. Omission of the hidden momentum is contrary to the position of standard modern electrodynamics texts, and leads to violation of Newton's law of action and reaction, negating Thomas's result.  Including the hidden momentum results in linear momentum conservation, but in the presence of Thomas precession, the total angular momentum is not generally conserved. The total angular momentum precesses for non-aligned spin and orbit, even in the absence of externally-applied magnetic field. As Thomas observes that secular angular momentum conservation is a necessary condition for a consistent simultaneous description of spin-orbit coupling and the anomalous Zeeman effect, such is not possible within classical electrodynamics in its absence.   
 
\end{abstract} 

% \begin{keywords}
%   Bohr Correspondence Principle, Spin-orbit Interaction, intrinsic angular momentum,
%   Larmor precession, Thomas precession, Thomas factor,
%   orbital angular momentum, quantum theory
% \end{keywords}

%\end{titlepage}

%\pagebreak
% PACS, the Physics and Astronomy Classification Scheme.
      
% 31.15Gy,       semiclassical methods (2006)
% 41.20.-q     Applied classical electromagnetism
% 45.05+x  General theory of classical mechanics of discrete systems 
% 45.20.-d Formalisms in classical mechanics  
% 45.20.D- Newtonian mechanics  
% 45.20.da Forces and torques  
% 45.20.dc Rotational dynamics  
% 45.20.df Momentum conservation  

               % 
%\twocolumn
%\addcontentsline{hydrogen}{section}{Introduction}
% 45+

\maketitle

%\begin{widetext}
%\pagebreak
%\tableofcontents
%\pagebreak
%\end{widetext}

\section{Introduction}

L. H. Thomas (1903-1992), in his landmark paper of 1927 \cite{Thomas1927}, explained the then-anomalous factor of one-half in atomic spin-orbit coupling as due to the relativistic effect now known as Thomas precession. The Thomas precession reduces the Larmor precession angular velocity of the electron magnetic moment by one half.  Thomas's explanation required principally, in addition to special relativity and the kinematic effect that now bears his name, that the total of the secular, or orbit-averaged, angular momenta of the electron orbit and spin be a conserved quantity.  He concluded, ``The physical interest of the result obtained is that it shows [....] the correct doublet separation at the same time as the anomalous Zeeman effect [....]. These explanations do not seem to require anything more of the extra terms in the equations of motion of the electron than that its axis should precess about a magnetic field \(\mbox{\boldmath$H$}\) with angular velocity \( (e/mc) \mbox{\boldmath$H$} \), that in revolution in an orbit there be some secularly conserved angular momentum, and that the contribution of the electron to this angular momentum be \(h/4\pi\).''  

In addition to obtaining the correct spin-orbit coupling simultaneously with the anomalous Zeeman effect, Thomas's 1927 paper contains other important analyses. His equation of motion of the electron axis in an electromagnetic field is a forerunner of the covariant Bargmann-Michel-Telegdi (BMT) equation \cite{BMT1959} that provides a basis for measurement of the anomalous magnetic moment. Further, although as the title suggests his analysis is primarily kinematical, Thomas also touched on dynamical aspects of the spinning, magnetic electron's interaction with an atomic nucleus. In his analysis he made an assumption that the secular total electron orbital and intrinsic angular momentum is conserved.  As quoted above, this conservation of the total secular angular momentum was observed by him to be essential to the existence of a well-defined intrinsic angular momentum, as well as to its utility in explaining the then-anomalous spin-orbit coupling.  He also reported, without providing details, that secular angular momentum conservation was achieved dynamically for either of two possible electron equations of translational motion in the electromagnetic field.  The general conclusion that could be drawn, and that continues to hold sway to the present day, is that classical electrodynamics is quite successful in explaining the energetics of the spin-orbit interaction including the anomalous Zeeman effect.  Once the essentially quantum-mechanical features of the electron, \textit{i.e.}, its nonclassical gyromagnetic ratio and ``two-valuedness'', and wave nature, are accepted, then classical electrodynamics could accommodate the spinning electron without contradiction and quantitatively describe phenomena such as the doublet spacing. 

The present work argues that the situation is not so straightforward as proposed by Thomas.  Because one of Thomas's fundamental assumptions is not supported dynamically, his semiclassical description of the spin-orbit coupling, which still pervades the pedagogic literature, is cast into doubt. Furthermore, analysis using the magnetic dipole equation of translational motion according to the current textbooks also does not yield a simultaneous description of spin-orbit coupling and the anomalous Zeeman effect in accordance with empirical observation.  The proper semiclassical dynamical description is seen however to possess certain features that are generally regarded as exclusively quantum mechanical, and which are absent from Thomas's treatment.  In particular, the present analysis indicates the total angular momentum must precess even in the absence of an externally-applied magnetic field.

The present analysis indicates Thomas's conclusion as quoted above, although strictly correct, is not consistent with classical electrodynamics in its current understanding.  The derivation he presented was based on an assumption of secular angular momentum conservation rather than its determination from the electrodynamics of his model.  In order to electrodynamically obtain equations of motion where there is a secularly conserved angular momentum, it is necessary to assume a certain specific force on a magnetic moment in an anisotropic magnetic field.  Thomas reported examining separately two possible dynamical laws for the force on a magnetic electron in the anisotropic magnetic field of its rest frame, to determine whether the resulting dynamics conserved total secular angular momentum as required.  Although Thomas concluded that total secular angular momentum was conserved in both cases, provided an appropriate assumption is made regarding the gyromagnetic ratio of the electron, it is shown this is true only for the first case he considered.  The force law required to provide general secular mechanical angular momentum constancy in the absence of an externally-applied magnetic field, which was evaluated by Thomas and found by him to yield secular total angular momentum constancy, is not in agreement with the current textbook form \cite{griffiths,jcksn:classelec3} that would account for the ``hidden momentum'' of the electron intrinsic magnetic dipole moment in the Coulomb field of the proton. It is argued here that including the hidden momentum, as prescribed by the modern textbooks, leads to nonconservation of secular angular momentum when Thomas precession becomes significant. 

Although Thomas made an error that obscured the possibility secular angular momentum conservation was not assured, this error occurred only for an assumed translational force that was not in accordance with Thomas's electron model based on the prior work of Abraham, and so was of little consequence to the central results of his 1927 paper.  In the present day however the force law he analyzed as a second and less important case is taken to be the more correct one, and its acceptance as such leads to invalidation of Thomas's conclusion that secular angular momentum is conserved in the presence of his relativity precession.  Carrying the analysis based on the modern force law to its logical conclusion, it is also seen that a consistent picture cannot be constructed where the Thomas precession accounts for the anomalous factor of one-half in the spin-orbit coupling simultaneously with correctly describing the anomalous Zeeman effect.

\section{Outline of Thomas's paper and the present analysis of it}

Thomas's 1927 paper consists of nine sections titled as follows:

1. Notation.
2. Lorentz Transformations.
3. The kinematical description of the motion of an electron.
4. A first approximation to the rate of change of direction of the axis of the electron.
5. The Abraham spinning electron.
6. The secular change in the direction of the axis of an electron revolving in an orbit.
7. The application of the correspondence principle to obtain approximate term values.
8. A summary of the logical train by which the Goudsmit-Uhlenbeck theory connects the anomalous Zeeman effect with the optical and relativity doublets and accounts for them both as manifestations of the magnetic properties of the electron.
9. Note added later.

The interest of the present analysis is primarily on Thomas's Section 6.  It will be necessary  also to make note of certain assumptions and specializations made at the end of his Section 4 and as used in his Section 6.  Although Thomas's result will be seen to be valid, as he claimed, for an electron with a magnetic moment magnitude of the Bohr magneton and a gyromagnetic ratio of \(\lambda = e/(mc)\), for any dynamics that results in secular total angular momentum conservation, it will also be seen that obtaining secular angular momentum conservation is not so straightforward as Thomas concluded based on then-current understanding. Rather, it seems to be impossible within classical electrodynamics as it is understood today.   

The following summarizes the content of the present paper.  After introductory material and explanation of notation and units in Section III, Section IV provides an overview of Thomas's analysis that was based initially on his assumption of angular momentum secular conservation, leading to his equations of motion of the electron axis and the orbit normal. It is seen plainly that as observed by Thomas, these depend also on the electron being described by a nonclassical gyromagnetic ratio. 

Section IVA reexamines Thomas's conclusion that total angular momentum is a secular constant of the motion, in his atomic model, for either of two possible laws for the translational force on a magnetic dipole in an anisotropic magnetic field.  It is seen that the angular momentum is proven secularly conserved only for the first law Thomas considered.

Section IVB discusses, in light of the modern understanding of the hidden mechanical momentum of a magnetic dipole in an electric field, which force law is the correct one.  It is concluded that the correct force law is the one that was not proven to secularly conserve angular momentum in the presence of Thomas precession.  

Section V duplicates results by Thomas, and provides intermediate steps not found in the 1927 publication.  

VA.  The explicit derivation of Thomas's equation of motion of the electron axis in a central electric field and assuming a gyromagnetic ratio (in Gaussian units) of \(\lambda=e/(mc)\), or equivalently a g-factor, \(g=2\).         

VB. Explicit evaluation of the torque on the orbit and the resulting equation of secular motion of the orbit normal, according to his Equation (5.1).  Thomas observes that his Eq. (5.1) obtains secularly conserved angular momentum but does not include the details in his paper. The result obtained herein is in agreement with Thomas.

Section VI provides further analyses.

VIA.  Examination of an alternative equation of motion of the electron axis in a central electric field, also based on Thomas's general equation of motion, but assuming a unity g-factor.  By inspection it can be seen, in full agreement with Thomas, that this case will not either result in angular momentum secular conservation.  

VIB. Since the equation of translational motion of a magnetic dipole including hidden momentum has an additional term not in either of Thomas's, the relative magnitude of this term is evaluated to see if it is significant in the system under consideration.  It is seen to be of order \((v/c)^2\) compared to the included term and so inconsequential to the results of Thomas's paper.  

VIC. A hypothetical system similar to Thomas's model is constructed, but in which Thomas precession can have no effect, and for which secular angular momentum and linear momentum are both conserved using the modern form for the translational force on a magnetic dipole.

VID.  The force law that Thomas correctly observed leads to secular conservation of angular momentum in the presence of his relativity precession, is shown to result in linear momentum nonconservation.   

VIE.  It is shown, using the effective force on a magnetic dipole in an anisotropic magnetic field that accounts for the hidden momentum, that the total mechanical angular momentum is not a secular constant of the motion of Thomas's model.  

VII.  The time rate of change of the total angular momentum, including the hidden and field angular momenta, is evaluated.  It is found that angular momentum is nonconserved in an amount exactly equal to the motion of the spin angular momentum due to the Thomas precession.  Also, the need for orbit-averaging is obviated by inclusion of the hidden momentum.  Except from the effect of Thomas precession, the total angular momentum is an exact constant of the motion in the modern treatment. 

VIII. The implications of the foregoing analyses are discussed. It seems that neither Thomas's nor the present analysis supports that the Thomas precession can consistently account for the spin-orbit coupling anomaly simultaneously with the anomalous Zeeman effect. 

IX.  The nature of the motion of the total secular angular momentum is investigated.  It is found that the total secular angular momentum magnitude is a constant of the motion, and that even absent an external magnetic field it precesses around a fixed axis with constant angular velocity. In spite of the motion of the total angular momentum, magnetic dipole radiation is not produced provided that the gyromagnetic ratio is a specific value, approximately twice the classically-expected value.             

X.  The three major conclusions of the present paper are reiterated.
  
\section{Notation and Units}

\begin{table}
\caption{Summary of Symbols}
\begin{ruledtabular}
\begin{tabular}{lcc}
Quantity & Thomas & Alternatively \\
\hline
 \\ orbital angular momentum & \mbox{\boldmath$K$} & \mbox{\boldmath$L$} \\ spin angular momentum & \mbox{\boldmath$\Omega$}  & \mbox{\boldmath$s$} \\
spin angular velocity & \mbox{\boldmath$w$} &  -  \\ 
moment of inertia & \(I\) &  -  \\ 
intrinsic magnetic moment & \(l \mbox{\boldmath$w$}\) & \mbox{\boldmath$\mu$} \\ 
gyromagnetic factor & - & \(g\) \\ 
gyromagnetic ratio & \(\lambda\) & \(g e / (2 m c) \) \\ 
magnetic induction & \mbox{\boldmath$H$} & \mbox{\boldmath$B$} \\
total angular momentum & \mbox{\boldmath$J$} & \mbox{\boldmath$J$} \\
acceleration & \mbox{\boldmath$f$} & \mbox{\boldmath$a$} \\
relativistic factor & \(\beta\) & \(\gamma\) \\
atomic number & \(Z\) & - \\
electron charge & \(-e\) & \(-e\) \\
electron mass & \(m\) & \(m\),\(m_e\) \\
proton mass & - & \(m_p\) \\
proton-to-electron displacement & \mbox{\boldmath$r$} & \mbox{\boldmath$r$} \\
electron velocity & \mbox{\boldmath$v$} & \mbox{\boldmath$v$} \\
speed of light & \(c\) & \(c\) \\
reduced Planck's constant & \(h/(2\pi)\) & \(\hbar\) \\
proper time & \(s\) & - \\
\end{tabular}
\label{tab1}
\end{ruledtabular}
\end{table}

In order to facillitate direct comparison between the analyses here and the original work of Thomas, and yet maintain easy distinguishability between the present analyses and those of Thomas, two separate although largely redundant sets of symbols are used herein.  One set is that used by Thomas.  It is used to duplicate Thomas's original equations, and for  intermediate results that are implied but not provided explicitly by him, and consequences that follow directly from his.

The second symbol set is used for analyses that are distinct from those published by Thomas in his 1927 paper. These will include evaluation of a hypothetical system with intrinsic spin where Thomas precession can have no effect, to determine its momentum conservation properties and show that they are in accordance with expectations.
  
The notation as used by Thomas and the alternative are summarized and compared on Table 1.

Gaussian units are used throughout.

\section{Description of Thomas's analysis of secular angular momentum conservation, and the issue of what is the proper equation of the electron translational motion}

In this section Thomas's analysis of angular momentum secular conservation will be reviewed with reference to his equations, but without reproducing them in detail.  Readers may either follow along in the original work (currently available online at \href{http://home.tiscali.nl/physis/HistoricPaper/}{Physis Project}), or provisionally accept the statements herein about Thomas's work as accurate.  It is hoped that sufficient detail is provided so that at least the strucure of the argument may be grasped without direct reference to the Thomas paper.  However, should the analysis of this section seem insufficient, Section V here following will reproduce Thomas's in more detail, and in some cases fill in intermediate steps not published in the original.    

In his Section 6, Thomas proposed that the secular, \textit{i.e.}, orbit-averaged, total of the electron intrinsic and orbital angular momenta should be a constant of the motion.  With this assumption and that the electron gyromagnetic ratio is \(e/(mc)\), and a further assumption that the secular change in the orbit normal is proportional to the electron axis orientation, he proceeded to determine an equation of motion for the orbit, and the proportionality constant between the intrinsic angular momentum of the electron and the angular velocity of the orbit normal precession around it.  Thomas next states (and as will be confirmed in Section V herein) that this can be obtained alternatively based on an electron equation of translational motion he attributes to Abraham, which he has also noted previously has gyromagnetic ratio of \(\lambda=e/(mc)\), or equivalently a g-factor of two.  He next considers an alternative, ``expected'' force on a small magnet in a magnetic field, and concludes that it as well can result in secularly conserved angular momentum and a well-defined electron intrinsic angular momentum.  It is not difficult however to see that such is not the case, and that his second force expression leads instead to secular angular momentum nonconservation in the presence of his relativity precession.  Then, the question of which if either of his force laws is the correct one will be addressed, with the result to follow in Section VI herein that the effective force law according to the current textbooks does not yield angular momentum secular conservation.      

\subsection{Description of Thomas's finding of angular momentum secular conservation in the presence of the relativity precession of the electron axis}

Thomas states in his opening remarks as quoted above, that obtaining the correct value of the Zeeman splitting from spin-orbit interaction requires that the electron spin axis must precess around an external magnetic field \( \mbox{\boldmath$H$}\) with angular velocity \( (e/(mc))\mbox{\boldmath$H$} \).  A current-loop magnetic dipole will precess with angular velocity of half this value.  Since the rate of change of the angular momentum is equal to the torque, and the torque is proportional to the magnetic moment of the magnetic dipole, it is apparent that this rate of precession requires that the ratio of the electron intrinsic magnetic moment to its spin angular momentum be twice the classically-expected ratio.  In modern terminology, the g-factor, \(g\), is two rather than the classical value of unity. The gyromagnetic ratio, that is, the ratio of the magnetic moment to the angular momentum, may be written as \(\lambda = g e /(2 m c)\), so Thomas's statement that the gyromagnetic ratio must equal \( e /(m c)\) is equivalent to requiring that \(g=2\). 

In accordance with his opening remarks as quoted, Thomas obtains his equations of motion of the electron spin axis and orbit normal (his Eqs. (6.71) and (6.72)) based explicitly on the assumption that \(g=2\).  His Equation (4.122) for the motion of the spin axis  is derived from his more general Eq. (4.121) (equivalent to Jackson \cite{jcksn:ThomasEq} Eq. (11.170)) based solely on this assumption, and further specialization to the central Coulomb field of the nucleus obtains his Eq. (6.1).  Further specialization of his Eq. (6.1), based on an assumption that the angular velocity of the orbital  angular momentum is proportional to the electron intrinsic spin angular momentum vector, results in his Eq. (6.71).  Therefore, due to his explicit assumption that \(g=2\), the validity of his Eq. (6.71) is assured only for the nonclassical gyromagnetic ratio \(\lambda = e/(mc)\), and therefore, it is also necessary that his Eq. (6.72), his equation of motion of the orbit normal that depends on both the strength of the electron intrinsic magnetic dipole moment and on its intrinsic angular momentum magnitude, be consistent with \(\lambda = e/(mc)\).   

Thomas's Eqs. (6.71) and (6.72) are thus equations of secular motion of the spin and orbital angular momenta derived initially on the assumptions that \(g=2\), that there is a secularly conserved total angular momentum consisting of their sum, and that ``the change in direction of the orbit normal will be a rotation of the form''  

\begin{equation}
\frac{e}{2 m c}\mbox{\boldmath$H$} = \varpi \mbox{\boldmath$w$},
\label{ThomasEq6.5}
\end{equation}

which is Thomas's Eq. (6.5), and where the proportionality constant,  \(\varpi\), ``depends on the shape and size of the orbit only''.  Thomas shows, in the absence of an externally applied magnetic field, it follows that

\begin{equation}
\frac{d}{dt}\left(\mbox{\boldmath$K$} + \frac{K}{\sigma}\varpi \mbox{\boldmath$w$}\right) = 0,
\label{ThomasEqafterEq6.5}
\end{equation}

and so ``\(\mbox{\boldmath$K$} + (K/\sigma)\varpi \mbox{\boldmath$w$}\) is what is secularly unchanged.  If and only if  \((K/\sigma)\varpi\) is the same for all orbits, can this be divided into the angular momentum of the orbit and an angular momentum of the orbit to be attributed to the electron.'' Thomas's equations of motion for the secular changes of the spin and orbit, his Eqs. (6.71) and (6.72), then follow with the intrinsic angular momentum of the electron given by 
\(\mbox{\boldmath$\Omega$} = (K/\sigma)\varpi \mbox{\boldmath$w$}\).     
Thomas's derivation of his Eqs. (6.71) and (6.72) thus has not depended on any particular form for the non-Coulomb translational force on the electron and resulting torque on the orbit.  Rather it has depended on the existence of a secularly conserved total angular momentum and that the gyromagnetic ratio \(\lambda = e/(mc)\), \textit{i.e.}, that the g-factor equal 2.  

Thomas reports considering a specific electron model, attributed to Abraham, to determine if its electrodynamics result in secularly conserved total angular momentum.  Thomas's Section 5 summarizes the results of Abraham, stating that Abraham's model has \(\lambda = e/(mc)\) and translational force given by his Eq. (5.1) as 

\begin{equation}
m\mbox{\boldmath$f$} = -e \mbox{\boldmath$E$} +  l\nabla(\mbox{\boldmath$w$} \cdot \mbox{\boldmath$H$} ),
\label{Thomas5.1}
\end{equation}

where \(\mbox{\boldmath$f$}\) is the acceleration and \(l\) is defined in Abraham's model (according to Thomas) as

\begin{equation}
l = (-)\frac{a^2 e}{3 c}, 
\label{l_def}
\end{equation}

where \(a\) is the radius of the spherical shell of charge of the Abraham electron model. (The parenthetic minus sign is added here, to account for the electron charge being \(-e\), and as required to obtain an electron magnetic moment directed oppositely to the spin.) Thomas next states that carrying out the averaging of the torque over an orbit obtains his Eq. (6.72) with \(\mbox{\boldmath$\Omega$} = I \mbox{\boldmath$w$}\), where \(I\) is the moment of inertia given according to Thomas by 

\begin{equation}
I = \frac{2}{9}\frac{a e^2}{c^2}. 
\label{I_def}
\end{equation}

In Section VB herein the explicit averaging process is carried out with a result found to be fully in agreement with Thomas.  However the present analysis does not agree with what in the order of his paper is Thomas's next claim.  Specifically, the alternative equation of translational motion Thomas considers leads to a violation of his assumption of \(\lambda=e/(mc)\), so that his Eq. (6.71) no longer holds, and so it cannot be concluded that angular momentum is secularly conserved for his second case.  In section VIE herein it is shown  explicitly that this equation of motion leads to nonconservation of the secular total angular momentum, but to see the contradiction the detailed analysis is not required, nor is it necessary to assume the electron conform to the model of Thomas's Section 5.1.  Indeed, the next equation of translational motion considered by Thomas is not based on that model.  Thomas states, ``If \(m\mbox{\boldmath$f$}_0 = -e \mbox{\boldmath$E$}_0 +  (Ie/mc)(\mbox{\boldmath$w$} \cdot\nabla) \mbox{\boldmath$H$}_0\) is used instead of (5.1),[...] it would be natural to put \(2I \mbox{\boldmath$w$} = \mbox{\boldmath$\Omega$}\) and still obtain (6.71),(6.72).''  To see that \(2I \mbox{\boldmath$w$} = \mbox{\boldmath$\Omega$}\) is inconsistent with a g-factor of two, first recognize that the quantity \(Ie\mbox{\boldmath$w$}/mc\) is a magnetic moment. Then, if \(\mbox{\boldmath$\Omega$}\) is the electron intrinsic angular momentum, it must be relatable to the electron intrinsic magnetic moment \(\mbox{\boldmath$\mu$}\) as

\begin{equation}
\frac{I e}{m c}\mbox{\boldmath$w$} \equiv \mbox{\boldmath$\mu$} = \frac{g e }{2 m c} \mbox{\boldmath$\Omega$},
\label{Edue2pcharge}
\end{equation}

or

\begin{equation}
2 I\mbox{\boldmath$w$} = g \mbox{\boldmath$\Omega$}.
\label{Edue2pcharge}
\end{equation}

So, setting \(\mbox{\boldmath$\Omega$} = I\mbox{\boldmath$w$}\) is equivalent to \(g=2\), but \(\mbox{\boldmath$\Omega$} = 2 I\mbox{\boldmath$w$}\) is equivalent to a unity \(g\).  Thus, the need in his second force law case to set \(\mbox{\boldmath$\Omega$} = 2 I\mbox{\boldmath$w$}\) is inconsistent with the assumptions used to derive his Eq. (6.71).  

In Section  IVB herein it is seen that of the two force laws considered, only the second is consistent with modern understanding of the equation of motion of a magnetic dipole in an anisotropic magnetic field and time-varying electric field.  

Additionally, it should be considered what equation of motion of the spin axis results in the case of unity \(g\).  Although it has been shown that Thomas's argument is based on the value \(g=2\), this does not necessarily preclude it being true nonetheless for unity \(g\).  However in this case, as shown in Section VIA herein, the spin axis in the laboratory frame is stationary, but the orbit normal precesses, also precluding conservation of the total secular angular momentum. 

\subsection{Possible equations of translational motion of the electron considered by Thomas, and the modern form accounting for the hidden momentum of a magnetic dipole}

To this point, two equations of translational motion of the magnetic electron have been discussed, these being the two proposed by Thomas.  They are Equation (\ref{Thomas5.1}) here which is his Equation (5.1), and 

\begin{equation}
m\mbox{\boldmath$f$} = -e \mbox{\boldmath$E$} - (Ie/mc)(\mbox{\boldmath$w$} \cdot\nabla) \mbox{\boldmath$H$},
\label{Thomas_EoTM_Case2}
\end{equation}

which is not numbered by Thomas or above where it is quoted as used by him. The minus sign on the second term on the right is added here as it is needed for consistency with the convention that the electron charge is \(-e\). Since these two equations of motion obtain quite different results in terms of the angular momentum secular conservation result or lack thereof, it is crucial to determine which of them, if either, is correct.

The translational force on a magnetic dipole of moment, \(\mbox{\boldmath$m$}\), in a possibly anisotropic magnetic field, \(\mbox{\boldmath$B$}\), sometimes referred to as the Stern-Gerlach force, is given generally by \cite{jcksn:classelec3}

\begin{equation}
\mbox{\boldmath$F$} = \nabla(\mbox{\boldmath$m$}\cdot\mbox{\boldmath$B$}).
\label{Jackson_3_Eq5.69}
\end{equation}

In recent decades, however, it has become recognized that a naive application of the Stern-Gerlach force in the equation of motion of a magnetic dipole fails to conserve the total linear momentum of a system consisting of a magnet interacting with a point charge \cite{ShockleyJames1967}.  The ``hidden'' mechanical momentum of a magnetic dipole in an electric field must be accounted for in the equation of motion \cite{Coleman1968}. A magnetic dipole located in an electric field and with magnetic moment not aligned exactly parallel to the field will have associated a nonzero Poynting vector and so a nonzero field momentum.  The hidden momentum is ``purely mechanical'' \cite{griffithsOC1} momentum equal and opposite the field momentum.  In relativistic dynamics properly the momentum changes in response to an applied force, and so this hidden momentum, given by \(\mbox{\boldmath$P$}_{\text{hidden}}=(\mbox{\boldmath$m$}\times\mbox{\boldmath$E$})/c\), must be incorporated into the equation of motion.  Although the force law is given by Jackson in all editions by Eq. (\ref{Jackson_3_Eq5.69}), we are instructed only in the third edition to incorporate the hidden momentum into the equation of translational motion of a magnetic dipole.  This results in an ``effective'' \cite{Hnizdo:1992} force law of

\begin{equation}
\mbox{\boldmath$F$} = \nabla(\mbox{\boldmath$m$}\cdot\mbox{\boldmath$B$}) - \frac{1}{c}\frac{d}{dt}(\mbox{\boldmath$m$}\times\mbox{\boldmath$E$}).
\label{HzizdoY91_Eq1}
\end{equation}

Now, as Thomas observes on his page 13, and as follows from vector calculus identities and Maxwell's equations,

\begin{equation}
(\mbox{\boldmath$m$}\cdot\nabla)\mbox{\boldmath$B$} - \nabla(\mbox{\boldmath$m$}\cdot\mbox{\boldmath$B$}) = -\mbox{\boldmath$m$}\times (\nabla \times \mbox{\boldmath$B$}) = -\frac{1}{c}(\mbox{\boldmath$m$}\times\dot{\mbox{\boldmath$E$}}),
\label{Identity}
\nonumber
\end{equation}

where the overdot indicates differentiation with respect to time.  Thus

\begin{equation}
 \nabla(\mbox{\boldmath$m$}\cdot\mbox{\boldmath$B$}) = (\mbox{\boldmath$m$}\cdot\nabla)\mbox{\boldmath$B$} + \frac{1}{c}(\mbox{\boldmath$m$}\times\dot{\mbox{\boldmath$E$}}).
\label{nabla_m_dot_B}
\end{equation}

Hnizdo's effective force becomes

\begin{equation}
\mbox{\boldmath$F$} = (\mbox{\boldmath$m$}\cdot\nabla)\mbox{\boldmath$B$} - \frac{1}{c}(\dot{\mbox{\boldmath$m$}}\times\mbox{\boldmath$E$}).
\label{HzizdoY91_Eq1a}
\end{equation}

The equation of translational motion for an electron model described by its mass \(m\), charge of \(-e\), and intrinsic magnetic moment \(\mbox{\boldmath$\mu$}\) then becomes

\begin{equation}
m\mbox{\boldmath$f$} = -e \mbox{\boldmath$E$} +   (\mbox{\boldmath$\mu$}\cdot\nabla)\mbox{\boldmath$B$} - \frac{1}{c}(\dot{\mbox{\boldmath$\mu$}}\times\mbox{\boldmath$E$}).
\label{ElecEoTMwHM}
\end{equation}

This is similar to Thomas's second considered equation of translational motion, which here is Equation (\ref{Thomas_EoTM_Case2}), but with addition of the term involving \(\dot{\mbox{\boldmath$\mu$}}\). The relative magnitude of this term compared to the\((\mbox{\boldmath$\mu$}\cdot\nabla)\mbox{\boldmath$B$}  \) term is easily evaluated for hydrogen for a circular orbit of one atomic unit radius in Section VIB herein, and found to be almost five orders of magnitude smaller. This is of order \( (v/c)^2 \) and so will be considered negligible in the present analysis. In any case it makes no difference to the  finding that total angular momentum is not a secularly conserved quantity in classical electrodynamics of an electron with intrinsic angular momentum.  It will be shown, the  \(\dot{\mbox{\boldmath$\mu$}}\) term is too small in the system under consideration to have any possibility of canceling the angular momentum imbalance due to the Thomas precession.

\subsection{Summary of conclusions to this point}

Thomas finds that in spite of the relativity precession of the electron axis, total angular momentum is secularly conserved if the electron gyromagnetic ratio is twice the classically-expected value, and if the translational force on the electron is given by either his Eq. (5.1) (Equation (\ref{Thomas5.1}) here) or the force law given by Equation (\ref{Thomas_EoTM_Case2}) here.  However, the analysis to this point has supported that only the first force law considered by Thomas was successfully demonstrated to result in angular momentum secular conservation, and that only the second force law he considered is similar to the correct one.  The conclusion based on analyses presented to this point is thus that according to modern electrodynamics, secular conservation of angular momentum in the presence of Thomas precession was not demonstrated by Thomas. 

The more detailed analyses to follow show explicitly that the modern standard laws of electrodynamics including hidden momentum conserve linear and secular angular momentum when Thomas precession has no effect. Also it is shown that in the presence of Thomas precession of the electron axis and in the system analyzed by Thomas, under these same laws, the total angular momentum that was considered by Thomas is not generally a secular constant of the motion.

\section{Duplication of Thomas's Result of Angular Momentum Secular Conservation by His Model}

In this section, Thomas's result that total angular momentum is a secular constant of the motion, based on the equation of electron translational motion according to his Equation (5.1), is reproduced.  This serves to lend credibilty to further analyses in Section VI herein showing the result of obtaining angular momentum secular conservation is less general than claimed by Thomas.  Some other features of Thomas's calculations as are important to the further analyses are noted as well.

\subsection{Thomas's equation of motion of the spin in the hydrogenic atom in a magnetic field}

Thomas's Eq. (6.71) for the motion of the electron axis is a central result of his 1927 paper, in that the calculation of his Section 7 of the correct value of the spin-orbit coupling is dependent upon it.  It will be worthwhile to reproduce Thomas's derivation here, to exhibit explicitly that it depends on the gyromagnetic ratio \(\lambda = e/(mc)\), equivalent to \(g=2\).  This would not be particularly an issue, since it is fully expected that \(g\) will be 2 and Thomas explictly states that it is required, except for one detail.  As shown above in Section IVA, when considering the alternative force law which was also found to yield secular angular momentum conservation, it was required in order to satisfy his Eq. (6.72) to assume implicitly a unity \(g\), and thus invalidate his use also of his Eq. (6.71) in determining the total secular angular momentum.  

The interest of the present analysis begins near the conclusion of Thomas's Section 4.  In that section, Thomas obtained that the equation of motion of the electron axis is given by (in his notation, his Eq. (4.121))

\begin{widetext}

\begin{equation}
\frac{d \mbox{\boldmath$w$}}{ds} = \left[\left( \frac{e}{mc} + \beta\left(\lambda-\frac{e}{mc}\right)\right)\mbox{\boldmath$H$} + \frac{1-\beta}{v^2}\left(\lambda-\frac{e}{mc}\right)\left(\mbox{\boldmath$H$} \cdot \mbox{\boldmath$v$} \right)\mbox{\boldmath$v$} + \left(\frac{e}{mc^2}\frac{\beta^2}{1+\beta}-\frac{\lambda\beta}{c} \right)\left[\mbox{\boldmath$v$} \times \mbox{\boldmath$E$} \right]\right] \times \mbox{\boldmath$w$},   
\label{Thomas4.121}
\end{equation}

\end{widetext}

where \(\mbox{\boldmath$w$}\) is the electron angular velocity or axis orientation as desired, \(s\) here is the electron proper time, \( \beta \) here is the relativistic factor \((1-(v/c)^2)^{-1/2}\), usually now denoted by \(\gamma\), and \( \lambda\) is the gyromagnetic ratio of the electron.  Specializing to \( \lambda = e/(mc)\) obtains straightforwardly and without approximation Thomas's Eq. (4.122),

\begin{equation}
\frac{d \mbox{\boldmath$w$}}{ds} = \left[\left( \frac{e}{mc}  \right)\mbox{\boldmath$H$} -  \frac{e}{mc^2}\left(\frac{\beta}{1+\beta} \right)\left[\mbox{\boldmath$v$} \times \mbox{\boldmath$E$} \right]\right] \times \mbox{\boldmath$w$}.   
\label{Thomas6.71_pre3}
\end{equation}

The electric field at the electron due to the central Coulomb field is

\begin{equation}
\mbox{\boldmath$E$} = \frac{e Z \mbox{\boldmath$r$}}{r^3},
\label{Edue2pcharge}
\end{equation}

where \( \mbox{\boldmath$r$}\) is the vector from the nucleus towards the electron. Substituting into Thomas's Eq. (4.122) gives in the low-velocity approximation

\begin{equation}
\frac{d \mbox{\boldmath$w$}}{ds} = \left[\left( \frac{e}{mc}  \right)\mbox{\boldmath$H$} -  \frac{e^2}{mc^2}\frac{1}{2}\left[\mbox{\boldmath$v$} \times \frac{\mbox{\boldmath$r$}Z}{r^3} \right]\right] \times \mbox{\boldmath$w$},   
\label{Thomas6.71_pre2}
\end{equation}

which is Thomas's Eq. (6.1).  This can be written in terms of the orbital angular momentum \(\mbox{\boldmath$K$} = \mbox{\boldmath$r$} \times m\mbox{\boldmath$v$} \) as

\begin{equation}
\frac{d \mbox{\boldmath$w$}}{ds} = \left[\left( \frac{e}{mc}  \right)\mbox{\boldmath$H$} +   \frac{1}{2}\frac{e^2}{m^2c^2} \frac{Z}{r^3}\mbox{\boldmath$K$} \right] \times \mbox{\boldmath$w$}.   
\nonumber
\end{equation}

Assuming the variation in \( \mbox{\boldmath$K$}\) and \( \mbox{\boldmath$w$}\) are small over the duration of an orbit revolution, and using angle brackets to denote a time average over a revolution of the orbit,

\begin{equation}
\left \langle \frac{d \mbox{\boldmath$w$}}{ds}\right \rangle = \left[\left( \frac{e}{mc}  \right)\mbox{\boldmath$H$} +   \frac{1}{2}\frac{e^2}{m^2c^2}\left \langle \frac{Z}{r^3}\right \rangle \mbox{\boldmath$K$} \right] \times \mbox{\boldmath$w$},   
\label{Thomas6.71_pre}
\end{equation}

which is Thomas's Eq. (6.2), except for the omission of the exponent on \(m\) in the denominator of the second term in Thomas's published version.  This omission is of no consequence to the findings of either Thomas or the present analysis.

Thomas defines

\begin{equation}
\mbox{\boldmath$\sigma$} = \frac{1}{2} \frac{e^2}{m^2c^2}\left \langle \frac{Z}{r^3}\right \rangle \mbox{\boldmath$K$},
\label{Thomas_sigma}
\end{equation}

which ``is equal to the rate of precession of the perihelion of the orbit in its own plane due to the Sommerfeld relativity effect,'' but here an extra \(m\) factor  has been added in the denominator compared to Thomas's published version, due to its omission in the original, as noted above.  Thus Thomas's Eq. (6.71) for the motion of the spin axis is obtained as

\begin{equation}
\frac{d \mbox{\boldmath$\Omega$}}{dt} = \left[\left( \frac{e}{mc}\mbox{\boldmath$H$} + \frac{\sigma}{K}\mbox{\boldmath$K$} \right) \times \mbox{\boldmath$\Omega$}   \right],
\label{Thomas6.71}
\end{equation}

and is thus clearly based on the assumption that \(\lambda = e/(mc)\) or equivalently, \(g=2\).

\subsection{Thomas's equation of secular motion of the orbital angular momentum based on his force law attributed to Abraham, and demonstration that it secularly conserves angular momentum}

As described in Section IVA herein, Thomas derives his Eq. (6.72) for the secular motion of the orbit normal based on assumptions of angular momentum secular conservation and that the motion due to the electron intrinsic magnetic moment is a rotation proportional to the electron axis orientation.  Thomas states that his Eq. (6.72) may be alternatively obtained using the force law of his Eq. (5.1) credited to Abraham, with \(Iw = \Omega\).  While agreeing, it is worthwhile to examine the alternative derivation in detail as a basis for what follows.  The analysis is simplified somewhat by assuming a circular orbit, but this restriction is not in general required.  Thomas did not assume a circular orbit. Thomas's notation is used except the magnetic field is denoted by \(\mbox{\boldmath$B$}\) rather than \(\mbox{\boldmath$H$}\), and the relativistic factor \((1-(v/c)^2)^{-1/2}\) is denoted by \(\gamma\) rather than Thomas's \(\beta\). 
   
\subsubsection{Derivation of Thomas's equation of secular motion of the orbital angular momentum based on his force law attributed to Abraham}

Thomas's Eq. (5.1) is 

\begin{equation}
m\mbox{\boldmath$f$} = -e \mbox{\boldmath$E$} +  l\nabla(\mbox{\boldmath$w$} \cdot \mbox{\boldmath$B$} ),
\label{Thomas5.1a}
\end{equation}

where \( \mbox{\boldmath$f$}\) is the acceleration. In Abraham's model, according to Thomas, with \(a\) the radius of the sperical shell of charge of the model, \(m = 2e^2/(3ac^2)\), and with \(l\) and \(I\) given by Eqs. (\ref{l_def}) and (\ref{I_def}), \(l=-eI/(mc)\) and so Eq. (\ref{Thomas5.1a}) becomes

\begin{equation}
m\mbox{\boldmath$f$} = -e \mbox{\boldmath$E$} -  \frac{eI}{mc}\nabla(\mbox{\boldmath$w$} \cdot \mbox{\boldmath$B$} ).
\label{Thomas5.1b}
\end{equation}

Letting \(Iw = \Omega\) as Thomas prescribes thus obtains

\begin{equation}
m\mbox{\boldmath$f$} = -e \mbox{\boldmath$E$} -  \frac{e}{mc}\nabla(\mbox{\boldmath$\Omega$} \cdot \mbox{\boldmath$B$} ).
\label{Thomas5.1c}
\end{equation}

Now, Eq. (\ref{Thomas5.1a}) strictly applies only in a reference frame where the electron is at rest, thus the applicable magnetic field is that in the electron rest frame.  (Strictly it should also be an inertial frame but this difficulty will not be dealt with in the present work.) Since the electromagnetic field in the laboratory frame due to the proton charge is fully determined, the magnetic field at the electron in the electron rest frame may be determined through the proper relativistic transformation of the field quantities from the laboratory frame to the electron rest frame.  This transformation will consist of a boost to the electron velocity and a gyration around the electron with angular velocity equal to that of the Thomas precession.  The electromagnetic field in a boosted frame may be found via a Lorentz transformation from an inertial frame as \cite{JacksonEMLT}     

\begin{equation}
\mbox{\boldmath$E$}' = \gamma\left( \mbox{\boldmath$E$} +  \mbox{\boldmath$\beta$}\times \mbox{\boldmath$B$}  \right)  -  \frac{\gamma^2}{\gamma+1}\mbox{\boldmath$\beta$}\left(\mbox{\boldmath$\beta$} \cdot \mbox{\boldmath$E$} \right),
\label{Expandt1a}
\end{equation}

\begin{equation}
\mbox{\boldmath$B$}' = \gamma\left( \mbox{\boldmath$B$} -  \mbox{\boldmath$\beta$}\times \mbox{\boldmath$E$}  \right)  -  \frac{\gamma^2}{\gamma+1}\mbox{\boldmath$\beta$}\left(\mbox{\boldmath$\beta$} \cdot \mbox{\boldmath$B$} \right),
\label{Expandt1a}
\end{equation}

with \(\mbox{\boldmath$\beta$} = \mbox{\boldmath$v$}/c\) here, where \(\mbox{\boldmath$v$}\) is the electron velocity in the laboratory frame.

Neglecting the motion of the proton around the center of mass, and its intrinsic magnetic moment, the electromagnetic field of the proton in the laboratory frame is a purely electric, radially-directed field.  Then the electric and magnetic fields in the frame boosted to momentarily match the electron velocity are

\begin{equation}
\mbox{\boldmath$E$}' = \gamma \mbox{\boldmath$E$}  -  \frac{\gamma^2}{\gamma+1}\mbox{\boldmath$\beta$}\left(\mbox{\boldmath$\beta$} \cdot \mbox{\boldmath$E$} \right),
\label{EFieldERFrame}
\end{equation}

\begin{equation}
\mbox{\boldmath$B$}' = -\gamma\left(\mbox{\boldmath$\beta$}\times \mbox{\boldmath$E$}  \right) ,
\label{HFieldERFrame}
\end{equation}

The order \(\beta^2\) difference in the electric field in the momentarily comoving frame as given by Eq. (\ref{EFieldERFrame}) vanishes in the case of a circular orbit and neglecting delay.  The approximation that \(\gamma\) is unity is continued. Applicability of Eq. (\ref{Jackson_3_Eq5.69}) in the noninertial electron rest frame is assumed. Also, it is assumed that any effects on the electrodynamics due to the Thomas precession are negligible, other than the standard centrifugal and Coriolis forces of nonrelativistic classical mechanics.  This level of approximation is adequate to duplicate Thomas's result.  

Assuming no external field Thomas's Eq. (6.72) is confirmed based on Eq. (\ref{Thomas5.1a}) and the approximations noted above, as follows.

For any vectors \(\mbox{\boldmath$a$}\) and \(\mbox{\boldmath$b$}\), 

\begin{equation}
\nabla(\mbox{\boldmath$a$}\cdot\mbox{\boldmath$b$}) = (\mbox{\boldmath$a$}\cdot\nabla)\mbox{\boldmath$b$} + (\mbox{\boldmath$b$}\cdot\nabla)\mbox{\boldmath$a$} + \mbox{\boldmath$a$}\times(\nabla\times\mbox{\boldmath$b$}) + \mbox{\boldmath$b$}\times(\nabla\times\mbox{\boldmath$a$}).
\nonumber
\end{equation}

Thus, with \(\mbox{\boldmath$\mu$} \equiv -e\mbox{\boldmath$\Omega$}/(mc)\), 

\begin{equation}
\nabla(\mbox{\boldmath$\mu$}\cdot\mbox{\boldmath$B$}) = (\mbox{\boldmath$\mu$}\cdot\nabla)\mbox{\boldmath$B$} + \mbox{\boldmath$\mu$}\times(\nabla \times \mbox{\boldmath$B$}).
\nonumber
\end{equation}

Substituting \(\dot{\mbox{\boldmath$E$}}/c\) for the curl of \(\mbox{\boldmath$B$}\) in vacuum and substituting for \(\mbox{\boldmath$E$} \) here as the Coulomb field of the proton at the electron position obtains

\begin{equation}
\nabla(\mbox{\boldmath$\mu$}\cdot\mbox{\boldmath$B$}) = (\mbox{\boldmath$\mu$}\cdot\nabla)\mbox{\boldmath$B$} + \mbox{\boldmath$\mu$}\times\left(\frac{1}{c}\frac{\partial}{\partial t}\left(\frac{e\mbox{\boldmath$r$}}{r^3}\right)\right),
\nonumber
\end{equation}

where \(\mbox{\boldmath$r$}\) is again the vector from the proton to the electron.  Under the further assumption of a circular orbit so that \(r = |\mbox{\boldmath$r$}|\) is constant, 

\begin{equation}
\nabla(\mbox{\boldmath$\mu$}\cdot\mbox{\boldmath$B$}) = (\mbox{\boldmath$\mu$}\cdot\nabla)\mbox{\boldmath$B$} - \mbox{\boldmath$v$}\times\left(\frac{e\mbox{\boldmath$\mu$}}{cr^3}\right).
\label{nabla_mu_dot_H_final}
\end{equation}

Substituting the magnetic field in the electron rest frame as given by Eq. (\ref{HFieldERFrame}) and approximating \(\gamma\) as unity, the first term on the right of Equation (\ref{nabla_mu_dot_H_final}) becomes 

\begin{equation}
(\mbox{\boldmath$\mu$}\cdot\nabla)\mbox{\boldmath$B$} = -(\mbox{\boldmath$\mu$}\cdot\nabla)\left(\frac{e}{cr^3} \left(
\mbox{\boldmath$v$} \times \mbox{\boldmath$r$}   \right) \right),
\label{Expandt1}
\nonumber
\end{equation}

which evaluates to

\begin{equation}
(\mbox{\boldmath$\mu$}\cdot\nabla)\mbox{\boldmath$B$} = -\frac{e}{cr^3} \left(
\mbox{\boldmath$v$} \times \mbox{\boldmath$\mu$}   \right) + \frac{3 e (\mbox{\boldmath$\mu$}\cdot\mbox{\boldmath$r$})\left(
\mbox{\boldmath$v$} \times \mbox{\boldmath$r$}   \right)}{cr^5}, 
\label{mudotnablaH}
\end{equation}

and so

\begin{widetext}

\begin{equation}
\mbox{\boldmath$r$}\times\left[(\mbox{\boldmath$\mu$}\cdot\nabla)\mbox{\boldmath$B$}\right] = \mbox{\boldmath$r$}\times\left[\frac{e\mbox{\boldmath$v$}}{c} \times \left(\frac{3(\mbox{\boldmath$\mu$}\cdot\mbox{\boldmath$n$})
\mbox{\boldmath$n$} - \mbox{\boldmath$\mu$}}{r^3}\right)  \right],
\label{TorqueModern}
\end{equation}

where \(\mbox{\boldmath$n$}\equiv\mbox{\boldmath$r$}/r\).  Incorporating the result of Eq. (\ref{nabla_mu_dot_H_final}), the torque on the circular orbit based on the force law of Thomas's Eq. (5.1) attributed to Abraham may now be written as

\begin{equation}
\mbox{\boldmath$\tau$}_K \equiv \mbox{\boldmath$r$} \times m\mbox{\boldmath$f$} = \mbox{\boldmath$r$}\times\left[\nabla(\mbox{\boldmath$\mu$}\cdot\mbox{\boldmath$B$})\right] = \mbox{\boldmath$r$}\times\left[\frac{e\mbox{\boldmath$v$}}{c} \times \left(\frac{ 3  (\mbox{\boldmath$\mu$}\cdot\mbox{\boldmath$n$})
\mbox{\boldmath$n$}- 2 \mbox{\boldmath$\mu$}}{r^3}\right)  \right].
\label{TorqueThomas5.1}
\end{equation}

\end{widetext}

Before proceeding to determine the equation of the secular motion of the orbit normal based on Eq. (\ref{TorqueThomas5.1}), it is worthwhile to consider Eq. (\ref{TorqueModern}) relative to Eq. (\ref{TorqueThomas5.1}).  Equation (\ref{TorqueModern}) is the ``effective'' torque on the electron orbit in the electron rest frame according to the modern effective force law of Eq. (\ref{HzizdoY91_Eq1}) here, neglecting the \(\dot{\mbox{\boldmath$\mu$}}\) term that is shown to be of order \(\beta^2\) in Section VIB.  The torque of Equation (\ref{TorqueModern}) is effective in the sense that it is based on the effective force on a magnetic dipole, that accounts for the presence of the hidden momentum.  Thus, the effective torque is equal to the rate of change of the conventional orbital angular momentum that excludes the hidden orbital angular momentum.  An alternative but equivalent treatment would be to derive the torque from the force on a magnetic dipole according to Equation (\ref{Jackson_3_Eq5.69}), and equate it to the time rate of change of the total orbital angular momentum that includes the hidden orbital angular momentum.  The hidden orbital angular momentum will be seen later to make an important contribution to the total angular momentum, but for the present the analysis will parallel Thomas's and describe the secular motion of the orbital angular momentum that excludes the hidden orbital angular momentum. 

Also, the term on the right of Equation (\ref{TorqueModern}) in parentheses is formally identical to the magnetic field of a magnetic dipole of magnetic moment \( \mbox{\boldmath$\mu$}\) outside the source region.  This form for the force and torque on the electron orbit is necessary if the force on the proton in the electron rest frame is to be equal and opposite that on the electron, assuming that the magnetic field of a magnetic dipole can be carried over from an inertial frame to the noninertial electron rest frame without significant change of form from that in the inertial frame where it is at rest.  (That the Lorentz force carries over to noninertial frames is well-established, leaving only that the electromagnetic field due to the dipole must be approximately that of a magnetic dipole at rest in an inertial frame.) In the electron rest frame the proton is orbiting the electron, at a velocity approximately opposite the velocity of the electron in the laboratory frame, and so will experience a Biot-Savart force in transiting the magnetic field due to the electron intrinsic magnetic moment.  In order that the total Lorentz force on the proton be equal and opposite the force on the electron, in accordance with Newton's third law and so conserving linear momentum, the torque on the electron is expected to be of the form of Eq. (\ref{TorqueModern}) rather than that of Eq. (\ref{TorqueThomas5.1}).  Therefore, as will be shown and in agreement with Thomas, although the effective force law of Eq. (\ref{TorqueThomas5.1}) leads to angular momentum secular conservation when Thomas precession is not negligible, it does not lead to a consistent dynamical description in which linear momentum is conserved.        

Continuing from Eq. (\ref{TorqueThomas5.1}), the vector triple product involving \(\mbox{\boldmath$\mu$}\) can be expanded as

\begin{equation}
\mbox{\boldmath$r$} \times (\mbox{\boldmath$v$} \times \mbox{\boldmath$\mu$}) = (\mbox{\boldmath$r$} \cdot\mbox{\boldmath$\mu$})\mbox{\boldmath$v$} - (\mbox{\boldmath$r$} \cdot \mbox{\boldmath$v$})\mbox{\boldmath$\mu$} 
= r(\mbox{\boldmath$n$} \cdot\mbox{\boldmath$\mu$})\mbox{\boldmath$v$}, 
\nonumber
\end{equation}

since the velocity and position vectors are orthogonal for the circular orbit so that the second term in the center vanishes.  Similarly the vector triple product of Eq. (\ref{TorqueThomas5.1}) involving \(\mbox{\boldmath$n$} \)  yields that \( \mbox{\boldmath$r$} \times (\mbox{\boldmath$v$} \times \mbox{\boldmath$n$}) = r \mbox{\boldmath$v$} \) and so Eq. (\ref{TorqueThomas5.1}) becomes

\begin{equation}
\mbox{\boldmath$\tau$}_K = \frac{e}{cr^2}
(\mbox{\boldmath$n$} \cdot \mbox{\boldmath$\mu$}) \mbox{\boldmath$v$} .
\label{torque_on_K}
\end{equation}

So, for any spin orientation other than parallel to the orbital angular momentum vector, where the torque vanishes, the torque is time-varying during the orbit.  The secular change in the orbital angular momentum is equal to the average torque over the course of an orbit.  The spin and orbital angular momentum vectors precess very slowly compared to an orbital period, so it is reasonable to treat their relative orientation as fixed during an orbit.  Choose for the electron rest frame Cartesian coordinate axes with directions \(\hat{\mbox{\boldmath$x$}}, \hat{\mbox{\boldmath$y$}},\hat{\mbox{\boldmath$z$}} \) with origin at the electron and where the orbital angular momentum \( \mbox{\boldmath$K$} \) is in the \( \hat{\mbox{\boldmath$z$}} \) direction.  Supposing that in general the proton spin is not aligned with \(\mbox{\boldmath$K$}\), choose the \( \hat{\mbox{\boldmath$x$}}\) direction to be aligned with the projection of \( \mbox{\boldmath$\Omega$} \) into the orbital plane.  The time origin may also be selected so that the electron at \(t=0\) is in the \( \hat{\mbox{\boldmath$x$}}\) direction from the proton. Then over a time interval where the precession of the spin axis may be neglected, \( (\mbox{\boldmath$n$} \cdot \mbox{\boldmath$\mu$}) =  \mu_\perp \cos(\omega t) \), and Eq. (\ref{torque_on_K}) can be rewritten as

\begin{equation}
\mbox{\boldmath$\tau$}_K = \frac{e}{cr^2}\mu_\perp \cos(\omega t)  \mbox{\boldmath$v$} ,
\nonumber
\end{equation}

where \(\mu_{\perp}\) is the proton intrinsic magnetic moment component in the orbital plane and \( \omega \) here is the orbital frequency of the  electron.  Expanding the velocity obtains

\begin{equation}
\mbox{\boldmath$\tau$}_K = \frac{e\mu_\perp \cos(\omega t)}{cr^2}  v (-\sin(\omega t)\hat{\mbox{\boldmath$x$}}  +   \cos(\omega t)\hat{\mbox{\boldmath$y$}}).
\nonumber
\end{equation}

Integrating over an orbital period \( T = 2\pi/\omega\) and dividing by \(T\) to obtain the average, the \(x\) component with \( \sin(\omega t) \cos(\omega t) \) vanishes and the \( \cos^2(\omega t) \) factor on the \(y\) component contributes a factor of a half and so 

\begin{equation}
\langle \mbox{\boldmath$\tau$}_K \rangle = \frac{e\mu_\perp}{2cr^2}  \frac{e}{\sqrt{m R}}\hat{\mbox{\boldmath$y$}},
\nonumber
\end{equation}

where angle brackets indicate the average over a turn of the orbit.  But

\begin{equation}
\mu_\perp\hat{\mbox{\boldmath$y$}} = -\mu_\perp (\hat{\mbox{\boldmath$x$}} \times \hat{\mbox{\boldmath$K$}}) \equiv -\mbox{\boldmath$\mu$}_{\perp}\times \hat{\mbox{\boldmath$K$}},
\nonumber
\end{equation}

where \(\hat{\mbox{\boldmath$K$}}\) is a unit vector in the direction of \(\mbox{\boldmath$K$}\).  Thus

\begin{equation}
\langle \mbox{\boldmath$\tau$}_K \rangle = -\frac{e^2 }{2 c R^{5/2} \sqrt{m}}  \mbox{\boldmath$\mu$}_{\perp} \times \hat{\mbox{\boldmath$K$}}
\nonumber
\end{equation}

is the average torque on the proton orbit in the electron rest frame due to the proton motion through the electron intrinsic magnetic field.  Using \(K=e\sqrt{mr}\) for the magnitude of \(\mbox{\boldmath$K$}\) for the circular orbit \cite{HallidayResnick} this becomes

\begin{equation}
\langle \mbox{\boldmath$\tau$}_K \rangle = \langle\dot{\mbox{\boldmath$K$}}\rangle = \mbox{\boldmath$K$} \times \frac{e}{2 c r^{3} m} \mbox{\boldmath$\mu$}_{\perp}. 
\nonumber
\end{equation}

The perpendicular component notation may be dropped since perpendicularity is defined here relative to the orbital plane, and replacing the intrinsic magnetic moment with its equivalent in terms of spin, and with Thomas's assumption that the electron gyromagnetic ratio \(\lambda=e/(mc)\), obtains 

\begin{equation}
\langle\dot{\mbox{\boldmath$K$}}\rangle =   \frac{e^2}{2 m^2 c^2 r^{3} } \mbox{\boldmath$\Omega$} \times \mbox{\boldmath$K$},
\label{Thomas6.72noext}
\end{equation}

which, with \(\mbox{\boldmath$\sigma$}\) given by Eq. (\ref{Thomas_sigma}) here, is equivalent to Thomas's Eq. (6.72) for no external field.

It is worth noting, particularly in that it is necessary to analyses following, that although Eq. (\ref{Thomas6.72noext}) was developed in the electron rest frame, it is also true in the laboratory frame.  In general the rate of change of any vector \(\mbox{\boldmath$G$}\) transforms from a rotating to an inertial laboratory frame as \cite{gldstn:classmech}

\begin{equation}
\left(\frac{d \mbox{\boldmath$G$}}{dt}\right)_{\text{lab}} = \left(\frac{d \mbox{\boldmath$G$}}{dt}\right)_{\text{rot}} +  \mbox{\boldmath$G$}  \times \mbox{\boldmath$\omega$},
\label{transf2lab}
\end{equation}

where \(\mbox{\boldmath$\omega$}\) is the angular velocity of the rotating frame. Here the rotation is due to the Thomas precession with angular velocity given by \cite{jcksn:ThomasEq}

\begin{equation}
\mbox{\boldmath$\omega$}_{\text{T}} = \frac{\gamma^2}{\gamma+1} \frac{\mbox{\boldmath$a$}  \times \mbox{\boldmath$v$}}{c^2}.
\label{TPAngVel}
\end{equation}

Neglecting delay the acceleration due to the Coulomb attraction is purely radial, so the Thomas precession angular velocity is parallel to the orbital angular momentum vector. The second term on the right of Eq. (\ref{transf2lab}) thus vanishes for \(\mbox{\boldmath$G$} \equiv \mbox{\boldmath$K$}\).  Also, the acceleration of the electron toward the proton being radial, the inertial force due to it also does not enter into the torque on the orbit, as it vanishes under the cross multiplication by \(\mbox{\boldmath$r$}\). The rate of change of the orbital angular momentum is thus the same in both the laboratory frame and the electron rest frame.

\subsubsection{Demonstration that Thomas's force law attributed to Abraham secularly conserves angular momentum in the presence of Thomas precession}

Substituting for \(\mbox{\boldmath$\sigma$}\) as given by Eq. (\ref{Thomas_sigma}) here, into Thomas's Eq. (6.71) (Eq. (\ref{Thomas6.71}) here) for the motion of the spin axis, obtains in the absence of an external field, and for hydrogen (\(Z=1\)) in the approximation of a stationary proton and assuming a circular electron orbit 

\begin{equation}
\frac{d \mbox{\boldmath$\Omega$}}{dt} = \frac{e^2}{2m^2c^2 r^3} \mbox{\boldmath$K$} \times \mbox{\boldmath$\Omega$}.
\label{Thomas6.71_3}
\end{equation}

Letting \( \mbox{\boldmath$J$} = \mbox{\boldmath$\Omega$}+ \mbox{\boldmath$K$}\) be the total angular momentum,  requiring that the total angular momentum be a secular constant of the motion may be expressed as

\begin{equation}
\left \langle\frac{d\mbox{\boldmath$J$}}{dt} \right \rangle = \langle\dot{\mbox{\boldmath$\Omega$}}\rangle + \langle\dot{\mbox{\boldmath$K$}}\rangle = 0,
\nonumber
\end{equation}

or, from Eqs. (\ref{Thomas6.72noext}) and (\ref{Thomas6.71_3})

\begin{equation}
\left[ \frac{e^2}{2m^2c^2 r^3} -  \frac{e^2}{2 m^2 c^2 r^{3} } \right] \mbox{\boldmath$K$} \times \mbox{\boldmath$\Omega$} = 0 .  
\label{torqueporbdtemm15}
\end{equation}

Since the leading factor vanishes generally the total angular momentum is a secular constant of the motion.

\begin{figure}
	\centering
		\includegraphics[width=0.4\textwidth]{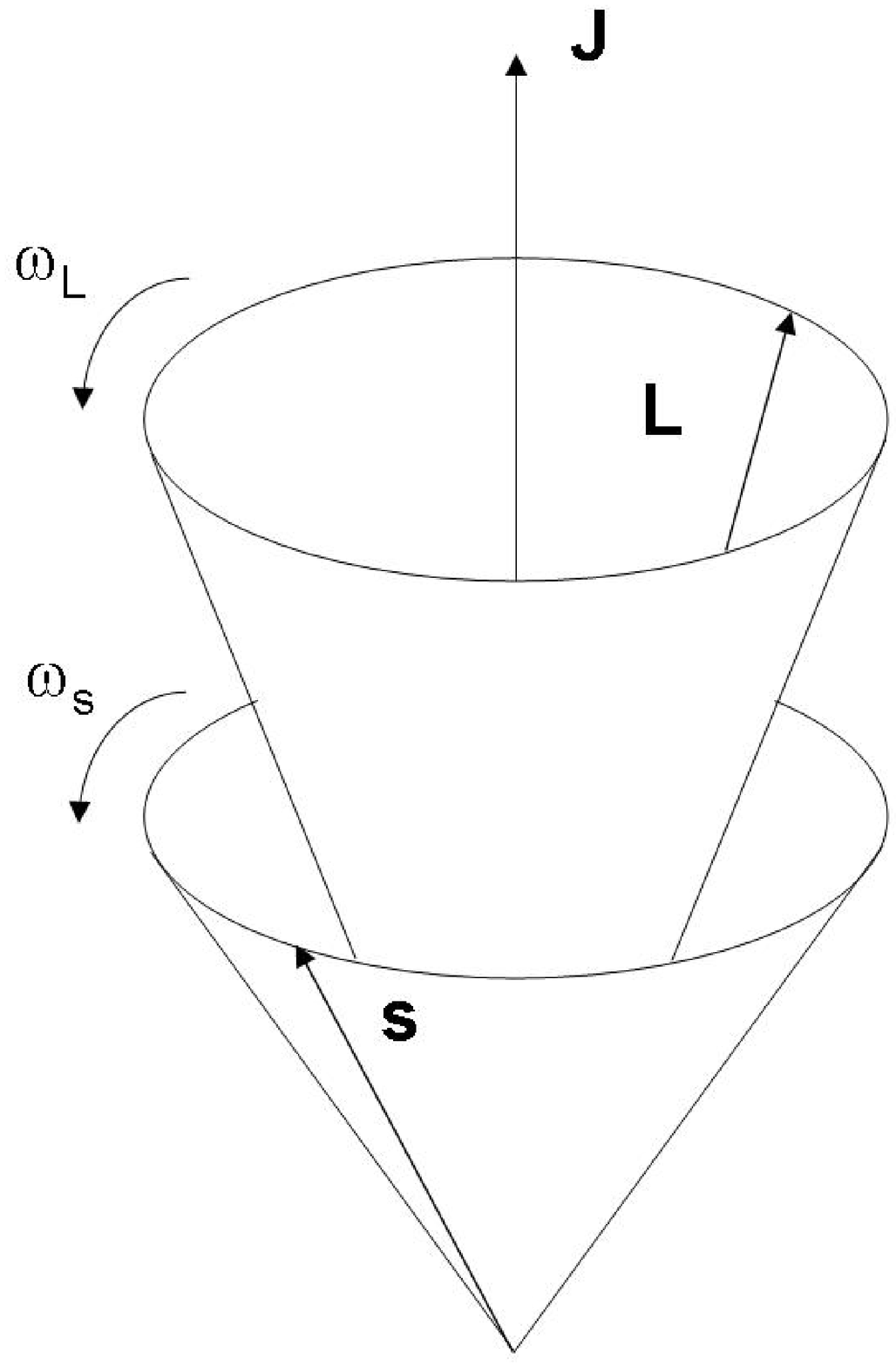}
	\caption{If \(\mbox{\boldmath$L$}\) and \(\mbox{\boldmath$s$}\) precess around \(\mbox{\boldmath$J$}\), constancy of \(\mbox{\boldmath$J$}\) requires that \(\mbox{\boldmath$L$}\) and \(\mbox{\boldmath$s$}\) precess with equal angular velocity. (After \protect\cite{Anderson:ModPhys}).}
	\label{fig:L_s_around_J}
\end{figure}

Thomas in his Section 7 provides a method for visualizing this situation that is illustrated on Figure 1 (with \(\mbox{\boldmath$K$} \equiv \mbox{\boldmath$L$}\) and \(\mbox{\boldmath$\Omega$} \equiv \mbox{\boldmath$s$}\)). Taking, for example, Eq. (\ref{Thomas6.71_3}), it may be understood \cite{GoldsteinOCAngVelDef} as the secular precession of the orbit normal around the electron axis with angular velocity given by  

\begin{equation}
\mbox{\boldmath$\omega$}_{\Omega} = \frac{e^2}{2m^2c^2 r^3}  \mbox{\boldmath$K$} = \frac{e^2 K}{2m^2c^2 r^3}  \hat{\mbox{\boldmath$K$}} , 
\label{KAngVel_1}
\end{equation}

where \(\hat{\mbox{\boldmath$K$}}\) is a unit vector in the direction of \( \mbox{\boldmath$K$}\).  This may be equally viewed as a secular precession of the orbit normal around the total angular momentum, since it is also true from Eq. (\ref{Thomas6.71_3}) that

\begin{equation}
\frac{d \mbox{\boldmath$\Omega$}}{dt} = -\frac{e^2}{2m^2c^2 r^3} \mbox{\boldmath$\Omega$} \times \left[ \mbox{\boldmath$\Omega$} +  \mbox{\boldmath$K$}\right] = -\frac{e^2}{2m^2c^2 r^3} \mbox{\boldmath$\Omega$} \times \mbox{\boldmath$J$}. 
\label{Thomas6.71_4}
\end{equation}

The angular velocity \(\mbox{\boldmath$\omega$}_{\Omega_J}\) of the electron axis precession around the total angular momentum pseudovector is then

\begin{equation}
\mbox{\boldmath$\omega$}_{\Omega_J} = \frac{e^2}{2m^2c^2 r^3}  \mbox{\boldmath$J$} = \frac{e^2 J}{2m^2c^2 r^3}  \hat{\mbox{\boldmath$J$}},  
\label{OmegaAngVelAroundTot}
\end{equation}

with \(J\equiv|\mbox{\boldmath$\Omega$}+\mbox{\boldmath$K$}|\).

Similarly, the secular angular velocity of the orbit normal around the total angular momentum is from Eq. (\ref{Thomas6.72noext}) 

\begin{equation}
\mbox{\boldmath$\omega$}_{K} = \frac{e^2}{2m^2c^2 r^3}  \mbox{\boldmath$\Omega$} = \frac{e^2 \Omega}{2m^2c^2 r^3}  \hat{\mbox{\boldmath$\Omega$}},  
\label{KAngVelAroundOmega}
\end{equation}

and the secular angular velocity of the orbit normal around the total angular momentum is 

\begin{equation}
\mbox{\boldmath$\omega$}_{K_J} = \frac{e^2}{2m^2c^2 r^3}  \mbox{\boldmath$J$} = \frac{e^2 J}{2m^2c^2 r^3}  \hat{\mbox{\boldmath$J$}} . 
\label{KAngVelAroundTot}
\end{equation}

Since by inspection of Eqs. (\ref{OmegaAngVelAroundTot}) and (\ref{KAngVelAroundTot}) the spin and orbit precession frequencies around the total are equal, the total secular angular momentum is a constant of the motion. This result is in full agreement with Thomas.

\section{Further Analyses}

Although no issue is taken here with respect to Thomas's conclusion that secular total angular momentum may be a constant of the motion given his equation of translational motion (5.1) (Equation (\ref{Thomas5.1}) here), it is worthwhile to examine this case in further detail.  It has been seen that considering the torque on the orbit according to Thomas's (5.1), the total secular angular momentum may be a constant of the motion. However, considering the opposition force on the proton due to the moving electron magnetic moment (or equivalently, in the electron rest frame, to the motion of the proton through the electron's intrinsic magnetic field), it is seen that Newton's third law of motion is not satisfied by Thomas's Eq. (5.1). 

Similarly, it is of interest to determine the momentum conservation properties of what according to modern textbooks is the correct equation of translational motion of the magnetic electron. It will be seen that accounting for hidden momentum, as expected, conserves linear momentum, and in the absence of Thomas precession conserves secular angular momentum. However in the presence of Thomas precession the secular angular momentum is not conserved. 

\subsection{Case of Unity \(g\)}

Given that Thomas's proposal, in the case of the second non-Coulomb translational force law he considered, to let the electron spin \(\Omega\) equal \(2 I w\) is as shown in Section IVA equivalent to a unity g-factor, it is here reconsidered whether total angular momentum can be secularly conserved in this case.  It may be borne in mind, this is not Thomas's intent, as he explicitly states in his opening a requirement that \(\lambda=e/(mc)\), equivalent to \(g=2\). In any case it is easily seen that secular angular momentum conservation does not result either in this case. 

It will be convenient to use a more modern symbology for this analysis. The notation of Jackson is used.   Thomas's Eq. (4.121) is translated into more modern form in terms of the g-factor, \(g\), replacing \(\beta\) by \(\gamma\), and with the spin angular momentum represented by \(\mbox{\boldmath$s$}\) rather than \(\mbox{\boldmath$\Omega$}\), by Jackson (except that here, as throughout, Thomas's convention that the electron charge is \(-e\) is used, rather than Jackson's, where \(e\) includes the charge polarity) as  

\begin{widetext}

\begin{equation}
\frac{d\mbox{\boldmath$s$}}{dt} = -\frac{e}{mc}\mbox{\boldmath$s$} \times \left[\left(\frac{g}{2}-1+\frac{1}{\gamma} \right)\mbox{\boldmath$B$} - \left(\frac{g}{2}-1\right)\frac{\gamma}{\gamma+1}(\mbox{\boldmath$\beta$} \cdot \mbox{\boldmath$B$})\mbox{\boldmath$\beta$} -  \left(\frac{g}{2}-\frac{\gamma}{\gamma+1}\right)\mbox{\boldmath$\beta$} \times \mbox{\boldmath$E$}\right].
\label{JacksonThomasEq}
\end{equation}

\end{widetext}

Specializing to \(g = 1\), noting that the magnetic field due to a nucleus possessing no magnetic moment vanishes in the laboratory frame, and approximating \(\gamma\) as unity, obtains that \(d\mbox{\boldmath$s$}/dt = 0\).  Thus, a classical current-loop magnetic dipole with net charge does not precess while nonrelativistically traversing an electric field, central Coulomb field included.  The Thomas precession cancels the Larmor precession when \(g\) is unity.  However for the nonrelativistic electron orbiting around a central charge, unless the orbit normal and spin axis are aligned, the orbit will nonetheless precess, and so the total mechanical angular momentum will not be even a secular constant of the motion.

This result is fully consistent with Thomas's observation that the correct simultaneous description of the spin-orbit coupling and anomalous Zeeman effect requires an electron gyromagnetic ratio \(\lambda=e/(mc)\).

\subsection{The magnitude of the difference between the translational force on the electron according to current textbooks and Thomas's ``expected'' force}

The proper form for the electron equation of translational motion, given by Eq. (\ref{HzizdoY91_Eq1}), differs from the second case considered by Thomas by the addition of the term involving \(\dot{\mbox{\boldmath$m$}}\).  In this section the relative magnitude of this term will be assessed for the case of an electron orbiting a proton circularly at the ground-state radius of the Bohr model.  

The force on the electron in its rest frame is given explicitly in terms of the electron intrinsic magnetic moment \(\mbox{\boldmath$\mu$}\) by Eq. (\ref{ElecEoTMwHM}) herein. For the electric Coulomb field of the proton, the torque on the orbit due to just the \(\dot{\mbox{\boldmath$\mu$}}\) term is then given by

\begin{equation}
\mbox{\boldmath$r$} \times \left[- \frac{1}{c}(\dot{\mbox{\boldmath$\mu$}}\times\mbox{\boldmath$E$})\right] = - \frac{ge^2}{m c^2r^3}\mbox{\boldmath$r$} \times (\dot{\mbox{\boldmath$s$}}\times\mbox{\boldmath$r$}) \equiv \mbox{\boldmath$\tau$}_{\dot{\mu}},
\label{HzizdoY91_Eq1_c1}
\end{equation}

where \(\mbox{\boldmath$r$}\) here and throughout is the vector from the proton to the electron, and \(r=|\mbox{\boldmath$r$}|\).

Translating Thomas's Eq. (6.71) (Eq. (\ref{Thomas6.71}) here) into the alternative notation, substituting for \(\sigma\) using Eq. (\ref{Thomas_sigma}), specializing to \(Z=1\) and a circular orbit, and with no externally-applied field obtains for the motion of the electron spin angular momentum

\begin{equation}
\dot{\mbox{\boldmath$s$}} =  \frac{\sigma}{L}\mbox{\boldmath$L$} \times \mbox{\boldmath$s$} =  \left(\frac{e^2}{2m^2c^2r^3} \right)\mbox{\boldmath$L$} \times \mbox{\boldmath$s$}.
\label{sdotfromThomas6.71}
\end{equation}

Substituting this for \(\dot{\mbox{\boldmath$s$}}\) in Equation (\ref{HzizdoY91_Eq1_c1})  yields for a circular orbit and with \(g=2\)

\begin{equation}
\mbox{\boldmath$\tau$}_{\dot{\mu}} =  \frac{e^4}{m^3 c^4r^4}(\mbox{\boldmath$n$} \times \left(\left(\mbox{\boldmath$s$} \times \mbox{\boldmath$L$} \right) \times\mbox{\boldmath$n$}\right)),
\label{HzizdoY91_Eq1_c2}
\nonumber
\end{equation}

or, at the Bohr radius, \(r_{\text{B}}\), where \(L = \hbar\) and with \(s=\hbar/2\),

\begin{equation}
\mbox{\boldmath$\tau$}_{\dot{\mu}} =  \frac{e^4 \hbar^2}{2 m^3 c^4{r_{\text{B}}}^4}\left(\mbox{\boldmath$n$} \times \left(\left(\hat{\mbox{\boldmath$s$}} \times \hat{\mbox{\boldmath$L$}} \right) \times\mbox{\boldmath$n$}\right)\right).
\label{HzizdoY91_Eq1_c2a}
\nonumber
\end{equation}

Since all of the vectors in the product on the right are of unit magnitude, the leading factor is an upper limit on the magnitude of the torque due to \(\dot{\mbox{\boldmath$\mu$}}\).  It is of interest to compare the magnitude of this torque with that of the torque on the orbit due to the other magnetic term of Eq. (\ref{ElecEoTMwHM}).  Letting \(x_{\text{B}}\) represent the ratio of the upper limit of the \(\dot{\mbox{\boldmath$\mu$}}\) term to the term involving \(\mbox{\boldmath$\mu$}\) at the Bohr radius, and using the upper limit of the \(\mbox{\boldmath$\mu$}\) term magnitude based on Eq. (\ref{sdotfromThomas6.71}) at the Bohr radius and with \(s=\hbar/2\)  obtains

\begin{equation}
x_{\text{B}} =  \frac{e^4 \hbar^2}{2 m^3 c^4{r_{\text{B}}}^4} \left[  \frac{e^2 \hbar^2}{c^2 {r_{\text{B}}}^3 m^2} \right]^{-1},
\label{HzizdoY91_Eq1_c3}
\nonumber
\end{equation}

or, with \(r_{\text{B}} = \hbar^2/ (e^2 m)  \),

\begin{equation}
x_{\text{B}} = \frac{e^4}{2 c^2 \hbar^2} \approx 2.7 \times 10^{-5}.
\label{RatioValatRB}
\end{equation}

Both of the terms in the ratio \(x_{\text{B}}\) have the same dependence on the relative orientation of \(\mbox{\boldmath$L$}\) and \(\mbox{\boldmath$s$}\), and so this is an upper limit on the magnitude of the \(\dot{\mbox{\boldmath$\mu$}}\) torque compared to the \(\mbox{\boldmath$\mu$}\) torque.  The error introduced by ignoring the \(\dot{\mbox{\boldmath$\mu$}}\) term is about the same magnitude as \((v/c)^2\) here. It will thus be neglected in the further analyses, except in Section VIC4, where it will be included in demonstrating that the effective force on a magnetic dipole that results through inclusion of the hidden momentum in the equation of translational motion, as embodied by Eq. (\ref{ElecEoTMwHM}), conserves linear momentum.

\subsection{Linear and secular angular momentum conservation of the modern force law when Thomas precession is negligible}

In order to construct a hypothetical bound system where the Thomas precession does not affect  the dynamics, a model described as follows will be used.  A heavy proton-like particle with spin and magnetic moment is bound to a light ``electron'' without spin or magnetic moment, and where the ``proton'' is sufficiently heavy compared to the electron so that the Thomas precession of its rest frame may be considered negligible.  In this system the Thomas precession has no significant effect on the dynamics of the spin and orbital angular momenta.  The electron rest frame may undergo Thomas precession, but for an electron without spin there will be no effect of the Thomas precession on the total angular momentum.  What then are the force laws and g-factor combinations that yield linear and secular angular momentum constancy? It is seen, incorporating the hidden momentum of a magnetic dipole in an electric field into the equation of motion, total angular momentum is a secular constant of the motion, independent of the value of \(g\) and for all orbital radii.  Linear momentum of the system is conserved as well.    

The non-magnetic electron orbits in the intrinsic magnetic field of the proton. 
The magnetic field at a point outside the source region is given in terms of the magnetic moment \(\mbox{\boldmath$m$}\) of the source as  \cite{jcksn:classelec_B_mm}  

\begin{equation}
\mbox{\boldmath$B$} = \frac{3\mbox{\boldmath$n$} \left(
\mbox{\boldmath$n$} \cdot \mbox{\boldmath$m$}   \right)  -
\mbox{\boldmath$m$}}{r^3},
\label{Bdue2MagMom}
\end{equation}

where \( \mbox{\boldmath$n$}=\mbox{\boldmath$r$}/r \) is a unit vector in the direction from the source to the field point (from proton to the electron here), and \( r \equiv |\mbox{\boldmath$r$}| \). 

The proton intrinsic magnetic moment will be  represented as

\begin{equation}
\mbox{\boldmath$\mu$}_p = \frac{g_p e}{2 m_p c} \mbox{\boldmath$s$}=
\frac{g_p e s}{2 m_p c}
\hat{\mbox{\boldmath$s$}},
\label{defofintrnsicmm}
\end{equation}

where \( s \) is the proton spin angular momentum magnitude  and
\(\hat{\mbox{\boldmath$s$}}\) is a unit-magnitude orientation
vector, \( m_p\) is the  proton mass, and \(g_p\) is the proton gyromagnetic factor.  In order to maintain that the magnetic moment of the proton is non-negligible in spite of it being arbitrarily massive so that its motion is negligible, it may be supposed that \(g_p\) is a large value here that grows in proportion to the proton mass.

\subsubsection{Secular motion of the orbit normal in the intrinsic magnetic field of a heavy  proton with spin}

The torque on the electron orbit around the proton is evaluated using Eq. (\ref{Bdue2MagMom}) as

\begin{equation}
\mbox{\boldmath$\tau$}_L = \mbox{\boldmath$r$} \times
\mbox{\boldmath$F$}_e = \mbox{\boldmath$r$} \times \left(
\frac{e}{c}\mbox{\boldmath$v$} \times \frac{3\mbox{\boldmath$n$}
\left( \mbox{\boldmath$n$} \cdot \mbox{\boldmath$\mu$}_p   \right) -
\mbox{\boldmath$\mu$}_p}{r^3} \right),
\label{torqueporbdtemm}
\end{equation}

with \( \mbox{\boldmath$n$} = \mbox{\boldmath$r$} / r \) here, and \( \mbox{\boldmath$v$}\) is the electron velocity as measured in the laboratory frame.  

Proceeding similarly as above in Section VB obtains for the torque on the orbit, under the simplifying but nonessential assumption of a circular orbit,

\begin{equation}
\mbox{\boldmath$\tau$}_L = \frac{2e}{cr^2}
(\mbox{\boldmath$n$} \cdot \mbox{\boldmath$\mu$}_p) \mbox{\boldmath$v$}.
\label{torqueporbdtemm3}
\end{equation}

Taking the average of the torque over an orbit and replacing the proton magnetic moment \(\mbox{\boldmath$\mu$}_p\) with its equivalent in terms of the spin as given by Eq. (\ref{defofintrnsicmm}) obtains that

\begin{equation}
\langle\dot{\mbox{\boldmath$L$}}\rangle = \mbox{\boldmath$L$} \times \frac{e}{c r^{3} m_e} \frac{g_p e s}{2 m_p c} \mbox{\boldmath$\hat{s}$} .
\nonumber
\end{equation}

The secular angular velocity of the orbit around the proton spin axis is then

\begin{equation}
\mbox{\boldmath$\omega$}_L = \frac{e^2}{c^2  r^{3} } \frac{g_p s}{2 m_p m_e} \mbox{\boldmath$\hat{s}$} .
\label{omegaL}
\end{equation}

\subsubsection{Motion of the proton spin axis in the magnetic field due to the orbiting electron}

The torque on the proton spin due to the orbiting electron charge is given by 

\begin{equation}
\mbox{\boldmath$\tau$}_s = \mbox{\boldmath$\mu$}_p \times
\mbox{\boldmath$B$} =  -\frac{g_p e}{2 m_p c} \mbox{\boldmath$s$} \times
\mbox{\boldmath$B$},
\label{torqueespindtporb}
\end{equation}

where \(\mbox{\boldmath$B$}\) here is the magnetic field at the proton due to the electron orbital motion around the proton. An adequate approximate expression for \(\mbox{\boldmath$B$}\) can be obtained from the Li\'enard-Wiechert fields \cite{JacksonLWFields} velocity magnetic field term, neglecting the propagation delay and in the low-velocity limit so that the relativistic factor \(\gamma\) and denominator term \(1-\mbox{\boldmath$\beta$} \cdot \mbox{\boldmath$n$}\) may both be approximated as unity. The acceleration fields may be considered negligible compared to the velocity fields at the atomic range scale. With \(\mbox{\boldmath$v$}\) the electron velocity in the laboratory frame and \(\mbox{\boldmath$r$}\) the vector from the electron to the proton, these approximations yield  

\begin{equation}
\mbox{\boldmath$B$} = \frac{e}{c R^3}\mbox{\boldmath$v$} \times \mbox{\boldmath$r$} = \frac{e}{c r^3 m_e}\mbox{\boldmath$L$}.
\label{approxBduetoe}
\end{equation}

Eq. (\ref{torqueespindtporb}) becomes

\begin{equation}
\mbox{\boldmath$\tau$}_s = -\frac{g_p e}{2 m_p c} \mbox{\boldmath$s$} \times
\left( \frac{e}{c r^3 m_e}\mbox{\boldmath$L$}  \right).
\label{muecrossBL2}
\nonumber
\end{equation}

Substituting for the magnitude of \(\mbox{\boldmath$L$}\) as \(L = e \sqrt{m r}\)  \cite{HallidayResnick} obtains

\begin{equation}
\mbox{\boldmath$\tau$}_s  = -\frac{g e^3  s}{2 c^2 {m_e}^{3/2} r^{5/2}} \hat{\mbox{\boldmath$s$}} \times
 \hat{\mbox{\boldmath$L$}} ,
\label{muecrossBL4}
\nonumber
\end{equation}

or

\begin{equation}
\dot{\mbox{\boldmath$s$}} = -\mbox{\boldmath$s$} \times
 \mbox{\boldmath$\omega$}_s = -\mbox{\boldmath$s$} \times \left( \frac{g_p e^3}
{2 c^2 m_p {m_e}^{1/2}r^{5/2}} \right) \hat{\mbox{\boldmath$L$}}.
\label{Ns5}
\nonumber
\end{equation}

The angular velocity of the precession of the proton spin around the orbit normal is thus

\begin{equation}
\mbox{\boldmath$\omega$}_s = \left( \frac{g_p e^3}
{2 c^2 m_p {m_e}^{1/2}r^{5/2}} \right) \hat{\mbox{\boldmath$L$}}.
\label{omegaslf}
\end{equation}

\subsubsection{Secular constancy of the total angular momentum when Thomas precession is negligible}

For secularly constant total angular momentum and in accordance with the model that spin and orbital angular momenta magnitudes are constant, it must be required that 

\begin{equation}
\langle\dot{\mbox{\boldmath$J$}}\rangle = \mbox{\boldmath$L$} \times \mbox{\boldmath$\omega$}_L +
\mbox{\boldmath$s$} \times \mbox{\boldmath$\omega$}_{s} = 0.
\label{totangeom}
\nonumber
\end{equation}

This may be rewritten using Eqs. (\ref{omegaL}) and (\ref{omegaslf})  as

\begin{equation}
\mbox{\boldmath$L$} \times \omega_L \hat{\mbox{\boldmath$s$}} +
\mbox{\boldmath$s$} \times \omega_s \hat{\mbox{\boldmath$L$}} = 0,
\label{constJcond}
\nonumber
\end{equation}

or

\begin{equation}
(L\omega_L -  s\omega_s)
\hat{\mbox{\boldmath$s$}} \times  \hat{\mbox{\boldmath$L$}} = 0.
\label{constJcond4}
\nonumber
\end{equation}

For non-aligned spin and orbital angular momenta, this leads to allowed orbital angular momentum given by

\begin{equation}
L = \frac{\omega_s}{\omega_L} s.
\label{constJcond6}
\nonumber
\end{equation}

Substituting for \( \omega_L\) and \(\omega_s\) from (\ref{omegaL}) and (\ref{omegaslf}) and reducing yields 

\begin{equation}
\frac{\omega_s}{\omega_L} s =  e \sqrt{mr}.
\label{conditiononL3}
\nonumber
\end{equation}

The right hand side is simply the angular momentum of the circular orbit. The total angular momentum averaged over a turn of the orbit is thus a constant of the motion for all orbital radii, and independent of the value of \(g\), in the hypothetical system where Thomas precession has no contribution.

\subsubsection{Conservation of the total linear momentum through inclusion of the hidden momentum}

Conservation of the linear momentum of the bound system of the heavy spinning proton and spinless electron requires that Newton's law of action and reaction hold.  In an inertial frame, it is necessary that the force on the proton be equal and opposite the force on the electron, or otherwise the center of mass will not remain unaccelerated.

As observed in Section IVB above, it is now generally recognized that the proper equation of motion of a magnetic dipole, \(\mbox{\boldmath$m$}\), in an electric field must account for the so-called hidden momentum, \(\mbox{\boldmath$m$} \times \mbox{\boldmath$E$}/c\). The hidden momentum is purely mechanical momentum that is evidenced by the nonvanishing Poynting vector of a dipole's magnetic field coexisting with a local electric field.  In the case of a charge moving in the field of a stationary magnetic dipole, it is straightforward to show that linear momentum conservation requires that the hidden momentum be incorporated in the equation of motion of the dipole.  

In the case of a nonstationary magnetic dipole, however, the demonstration that linear momentum is conserved is more difficult. The standard derivation \cite{JacksonVecPotMM} of the vector potential and electromagnetic field of a current-loop magnetic dipole assumes explicitly that the current distribution is stationary.  This poses a problem to the present analysis, in which the electromagnetic field of the (at least) precessing magnetic moment is needed to determine the exact force on the orbiting electron.  A rigorous determination in the literature of the electromagnetic field due to a nonstationary magnetic moment is not known to the author.  However, if it is assumed that the form of the vector potential due to a moving magnetic moment is formally identical to its form for the stationary moment, as it must be in the limit of vanishing motion, it is seen that linear momentum is conserved. (This approach to showing linear momentum conservation in the presence of a moving magnetic moment is the same as that taken by Hnizdo and other authors.) 

Due to the difficulty of obtaining rigorously an expression for the electromagnetic field of a precessing magnetic moment, the analysis of this section will proceed in two steps.  First, it will be shown that with inclusion of the hidden momentum, linear momentum is conserved for the case of a charge moving in the magnetic field of a stationary magnetic dipole.  This result is only in accordance with expectations based on current textbooks.  However it is worth presenting here for comparison with the alternative as used by Thomas, which will not obtain linear momentum conservation even neglecting any force arising from the motion of the proton magnetic moment, as is shown in Subsection D below. 

In the second step the effect of the precessional motion of the proton intrinsic magnetic moment is considered.  The force term due to the proton axis precession is available in Eq. (\ref{HzizdoY91_Eq1a}) for the force on the precessing dipole itself.  It is determined to be small in the regime of the present model, compared to the force already present from the existence of the dipole.  Then it is shown that the force on the magnetic dipole is equal and opposite the force on the charge provided that the vector potential of the magnetic dipole contains no additional terms due to nonstationarity of the magnetic moment.

For the first step consider the force on the proton intrinsic magnetic moment due to the anisotropy, at the proton, of the magnetic field due to the orbital motion of the electron charge.  Letting the magnetic moment \( \mbox{\boldmath$m$}\) of Eq. (\ref{HzizdoY91_Eq1a}) be the magnetic moment \(\mbox{\boldmath$\mu$}_p\) of the heavy proton, and substituting the right hand side of Eq. (\ref{mudotnablaH}) for the first term on the right of Eq. (\ref{HzizdoY91_Eq1a}) and with \(\mbox{\boldmath$B$}\) the magnetic field due to the motion of a charge, obtains that the total force on the proton intrinsic magnetic moment due to the circularly-orbiting electron charge is given by

\begin{equation}
\mbox{\boldmath$F$}_p = -\left(
\frac{e}{c}\mbox{\boldmath$v$} \times \frac{3\mbox{\boldmath$n$}
\left( \mbox{\boldmath$n$} \cdot \mbox{\boldmath$\mu$}_p   \right) -
\mbox{\boldmath$\mu$}_p}{r^3} \right)   - \frac{1}{c} \left(  \dot{\mbox{\boldmath$\mu$}}_p \times \mbox{\boldmath$E$},
\right)
\label{TotalForceOnProtMu}
\nonumber
\end{equation}

where \(\mbox{\boldmath$E$}\) here is the Coulomb field of the electron at the proton and so

\begin{equation}
\mbox{\boldmath$F$}_p = -\left(
\frac{e}{c}\mbox{\boldmath$v$} \times \frac{3\mbox{\boldmath$n$}
\left( \mbox{\boldmath$n$} \cdot \mbox{\boldmath$\mu$}_p   \right) -
\mbox{\boldmath$\mu$}_p}{r^3} \right)   - \frac{1}{c} \left(  \dot{\mbox{\boldmath$\mu$}}_p \times \frac{e\mbox{\boldmath$n$}}{r^2}
\right).
\label{Mu_p_dotForceOnProton}
\end{equation}

Assuming initially that the proton intrinsic magnetic moment is stationary, then the magnetic field at the electron is given by Eq. (\ref{Bdue2MagMom}), and so the force on the electron moving through this magnetic field is  

\begin{equation}
\mbox{\boldmath$F$}_e = \left(
\frac{e}{c}\mbox{\boldmath$v$} \times \frac{3\mbox{\boldmath$n$}
\left( \mbox{\boldmath$n$} \cdot \mbox{\boldmath$\mu$}_p   \right) -
\mbox{\boldmath$\mu$}_p}{r^3} \right) .
\label{torqueporbdtemm}
\nonumber
\end{equation}

So, the force on the electron in the intrinsic magnetic field of a proton with stationary magnetic moment is equal and opposite the force on the proton given by Eq. (\ref{Mu_p_dotForceOnProton}) for \(\dot{\mbox{\boldmath$\mu$}}_p = 0\).   Allowing for precessional motion of the proton magnetic moment, it may be noted again that in accordance with the analysis of Section VC, the second term on the right of Eq. (\ref{Mu_p_dotForceOnProton}) is small at atomic scales compared to the first.  Therefore, even were there no plausible explanation for a force opposite to the second term of Eq. (\ref{Mu_p_dotForceOnProton}), the linear momentum nonconservation due to this would be less than the linear momentum nonconservation of Thomas's assumed force as evaluated in Section VID following.  However, although there appears to be no rigorous solution in the literature to the question of what is the electromagnetic field due to a nonstationary magnetic dipole moment, it is clear that any such solution must reduce to the standard expression for a stationary magnetic moment in the limit of vanishing \(\dot{\mbox{\boldmath$\mu$}}_p\).  In the following any deviation of the vector potential from its form in the limiting case of the stationary magnetic moment will be assumed negligible for the magnitude of \(\dot{\mbox{\boldmath$\mu$}}_p\) expected in the current application.   

The vector potential {\boldmath $A$} due to a stationary magnetic dipole of moment
{\boldmath $m$} at a field point outside the source region, separated from {\boldmath $m$} by {\boldmath $r$} is \cite{JacksonVecPotMM} exactly

\begin{equation}
\mbox{\boldmath$A$} = \frac{\mbox{\boldmath$m$} \times
\mbox{\boldmath$r$}}{r^3}, \label{vecpotduetomm}
\end{equation}

and from the vector and scalar potentials the electric field is
given by

\begin{equation}
\mbox{\boldmath $E$} = - \frac{1}{c} \frac{ \partial \mbox{\boldmath
$A$}}{\partial t}  - \nabla \Phi  .
\nonumber
\end{equation}

The vector \(\mbox{\boldmath$r$}\) from the stationary proton to a field point is not time-varying.  Assuming that the form of the vector potential remains unchanged and that the scalar potential remains vanished for nonvanishing \(\dot{\mbox{\boldmath$m$}}\) thus obtains the electric field due to the orientational motion of the proton axis as 

\begin{equation}
\mbox{\boldmath $E$} = - \frac{1}{c} \frac{ \partial}{\partial t}\left( \frac{\mbox{\boldmath$m$} \times
\mbox{\boldmath$r$}}{r^3} \right) = - \frac{1}{c}\frac{\dot{\mbox{\boldmath$m$}} \times
\mbox{\boldmath$r$}}{r^3}.
\nonumber
\end{equation}

The total inertial-frame force on the spinless electron in the approximation of a stationary proton with nonstationary magnetic moment is then

\begin{equation}
\mbox{\boldmath$F$}_e = \left(
\frac{e}{c}\mbox{\boldmath$v$} \times \frac{3\mbox{\boldmath$n$}
\left( \mbox{\boldmath$n$} \cdot \mbox{\boldmath$\mu$}_p   \right) -
\mbox{\boldmath$\mu$}_p}{r^3} \right)  -   \frac{e \mbox{\boldmath$n$} \times \dot{\mbox{\boldmath$\mu$}}_p}{c r^2},
\label{Mu_p_dotForceOnElectron}
\end{equation}

which is equal and opposite the force on the proton as given by Eq. (\ref{Mu_p_dotForceOnProton}).  Inclusion of the hidden momentum in the equation of motion of the heavy magnetic proton has thus resulted in linear momentum conservation, as expected.

\subsection{Linear momentum nonconservation of Thomas's equation of translational motion of the electron}

As shown in Section IVB, Thomas's Eq. (5.1) (Eq. (\ref{Thomas5.1}) here) for the translational motion of the magnetic electron, does not incorporate the hidden momentum of the electron magnetic moment in the proton Coulomb field. It is to be expected therefore that his Eq. (5.1) will not result in linear momentum conservation by his model.  This nonconservation is demonstrated explicitly as follows.  

Starting with Eq. (\ref{nabla_m_dot_B}) for the general gradient of \(\mbox{\boldmath$m$}\cdot\mbox{\boldmath$B$}\), and evaluating the time derivative of the electric field due to circular orbital motion of the electron charge at the stationary heavy magnetic proton obtains 

\begin{equation}
\nabla(\mbox{\boldmath$\mu$}_p\cdot\mbox{\boldmath$B$}) = (\mbox{\boldmath$\mu$}_p\cdot\nabla)\mbox{\boldmath$B$} + \frac{e}{cr^3} \left(
\mbox{\boldmath$v$} \times \mbox{\boldmath$\mu$}_p   \right).
\label{GeneralGradOfMuDotB_2}
\end{equation}

Substituting for the first term on the right using Equation (\ref{mudotnablaH}) obtains

\begin{equation}
\nabla(\mbox{\boldmath$\mu$}_p\cdot\mbox{\boldmath$B$}) = -\frac{e}{c} \mbox{\boldmath$v$} \times \left( \frac{ 3 (\mbox{\boldmath$\mu$}_p\cdot\mbox{\boldmath$n$}) \mbox{\boldmath$n$} - 2 
 \mbox{\boldmath$\mu$}_p}{r^3}  \right).
\label{ThomasForceCircOrbit_2}
\end{equation}

This force is clearly not equal and opposite the force on the electron orbiting in the intrinsic magnetic field of the heavy proton.  The force on the orbiting electron must omit the factor of 2 on the second term in the numerator of the fraction in parentheses.  Furthermore, there is no question of the form of the force on the electron here, as there is no question of the magnetic field of a static magnetic moment. 

\subsection{Secular nonconservation of angular momentum of the modern effective force law in the presence of Thomas precession}

From Section VB1 herein, Eq. (\ref{TorqueModern}) is the torque on the electron orbit in the electron rest frame according to the effective force law of Eq. (\ref{HzizdoY91_Eq1}) here, that accounts for hidden momentum, neglecting the \(\dot{\mbox{\boldmath$\mu$}}\) term that was shown to be of order \(\beta^2\) in Section VIB.  Proceeding similarly to the analysis of Section VB1 based on Eq. (\ref{TorqueThomas5.1}), but starting with Eq. (\ref{TorqueModern}) rather than Eq. (\ref{TorqueThomas5.1}), arrives at 

\begin{equation}
\mbox{\boldmath$\tau$}_L = \frac{2e}{cr^2}
(\mbox{\boldmath$n$} \cdot \mbox{\boldmath$\mu$}) \mbox{\boldmath$v$} ,
\label{torque_e_orb}
\end{equation}

for the torque on the electron orbit.  This is exactly twice the torque of Eq. (\ref{torque_on_K}) based on the force law of Thomas's Eq. (5.1) (Eq. (\ref{Thomas5.1}) here).  The equivalent equation of motion to Thomas's Eq. (6.72) in the absence of an external magnetic field (Eq. (\ref{Thomas6.72noext}) here), but including hidden momentum and allowing for any g-factor value, is then

\begin{equation}
\langle\dot{\mbox{\boldmath$L$}}\rangle = \frac{ge^2}{2 m^2 c^2 r^{3} } \mbox{\boldmath$s$} \times \mbox{\boldmath$L$}.
\label{ModernLdot}
\end{equation}

The equation of motion for the spin in the laboratory frame, for the electron orbiting in the Coulomb field of the proton, may be obtained directly from Thomas's Eq. (4.121).  In the alternative notation and involving \(g\) explicitly this is Jackson's Eq. (11.170) (Eq. (\ref{JacksonThomasEq}) here). Specializing to no external fields, approximating \(\gamma\) as unity, and treating the proton as stationary obtains, analogously to Eq. (\ref{Thomas6.71_3}) here obtained from Thomas's (6.71) (Eq. (\ref{Thomas6.71}) here) but allowing for arbitrary g-factor,

\begin{equation}
\dot{ \mbox{\boldmath$s$}} = \left(\frac{g}{2} - \frac{1}{2} \right)\left(\frac{e^2}{m^2 c^2 r^3}\right) \mbox{\boldmath$L$} \times \mbox{\boldmath$s$}.
\label{SpinEOMAnyg}
\end{equation}

Requiring that the total angular momentum is secularly constant thus results in

\begin{equation}
\left \langle \frac{d \mbox{\boldmath$J$}}{dt} \right \rangle = \left(\frac{e^2}{m^2c^2 r^3}\right)\left[\frac{g}{2} - \left(\frac{g}{2} - \frac{1}{2} \right) \right] \mbox{\boldmath$s$} \times \mbox{\boldmath$L$} = 0  .
\label{ConstJanyg}
\nonumber
\end{equation}

If \(\mbox{\boldmath$s$} \times \mbox{\boldmath$L$} \neq 0 \), this requires that 

\begin{equation}
g=g-1,
\label{Contradiction}
\end{equation}

which is a contradiction for all finite values of \(g\).  The total angular momentum thus cannot be a secular constant of the motion in Thomas's model, when the hidden momentum is taken into account, for non-aligned spin and orbital angular momenta.

That Thomas precession is apparently inconsistent with conventional notions of angular momentum conservation has been noted previously by Phipps, who wrote \cite{Phipps:1986}, ``Whence cometh the energy of the Thomas precession, which by definition is of purely kinematic origin, all torques and physical energy sources being absent?''  Muller \cite{Muller:1992} however argues that Thomas precession is equivalent to an actual torque.  Since Muller's torque is internal and present in the absence of externally-applied magnetic field, it must lead to angular momentum nonconservation, or radiation. The issue of radiation due to the Thomas precession is addressed in Section IXC herein.

\subsection{Summary of results of the further analyses}

It has been seen that classical electrodynamics including the hidden momentum arrives easily at a solution that conserves both linear momentum and secular angular momentum for the system where Thomas precession is not relevant to the dynamics. However, assuming the same equation of translational motion of the magnetic electron applies in the presence of Thomas precession, leads to the impossibility of secular conservation of angular momentum by Thomas's model.          

%It seems unlikely that the laws of electrodynamics can be expected to be different %depending only on the presence or absence of Thomas precession.

\section{Angular momentum nonconservation due to the Thomas precession}

Up to this point the total angular momentum considered, as by Thomas, has consisted of the sum of the electron intrinsic spin angular momentum and the orbital angular momentum due to the motion of the electron mass orbiting around the proton.  In Thomas's analysis, this quantity is conserved only secularly.  The total angular momentum that is expected to be exactly conserved in the modern view consists of \cite{Hnizdo:TotLCons} the ordinary angular momentum as considered by Thomas, the hidden mechanical angular momentum given as \(\mbox{\boldmath$L$}_{\text{hidden}}  \equiv \mbox{\boldmath$r$} \times \mbox{\boldmath$P$}_{\text{hidden}}=\mbox{\boldmath$r$} \times \left(\mbox{\boldmath$\mu$} \times \mbox{\boldmath$E$}\right)/c\), and the field angular momentum around the proton. The field angular momentum in the present application, of the electron intrinsic magnetic moment in the proton Coulomb field, vanishes (see Trammel's  \cite{Trammel:1964} Equation (10) with \(\mbox{\boldmath$r$}_1 = \mbox{\boldmath$r$}\)), and so need not be included here. The total angular momentum then consists of    

\begin{equation}
\mbox{\boldmath$J$} = \mbox{\boldmath$L$} + \mbox{\boldmath$s$} + \mbox{\boldmath$L$}_{\text{hidden}},
\label{TotMechAngMom}
\end{equation}

with

\begin{equation}
\mbox{\boldmath$L$}_{\text{hidden}}  =  \frac{1}{c}\mbox{\boldmath$r$} \times \left(\mbox{\boldmath$\mu$} \times \mbox{\boldmath$E$}\right)  =  \frac{e}{c r^3}\mbox{\boldmath$r$} \times \left(\mbox{\boldmath$\mu$} \times \mbox{\boldmath$r$}\right). 
\label{L_hidden_def}
\end{equation}

The equation of motion of the total angular momentum including the hidden angular momentum is thus

\begin{equation}
\dot{\mbox{\boldmath$J$}} = \dot{\mbox{\boldmath$L$}} + \dot{\mbox{\boldmath$s$}} + \dot{\mbox{\boldmath$L$}}_{\text{hidden}},
\label{MotionTotMechAngMomWithHidden}
\end{equation}

To evaluate \(\dot{\mbox{\boldmath$L$}}\), consider that the torque on the orbit, \(\mbox{\boldmath$\tau$}_L\), can be evaluated in the electon rest frame using Equation (\ref{HzizdoY91_Eq1a}) to obtain

\begin{equation}
\dot{\mbox{\boldmath$L$}} = \mbox{\boldmath$\tau$}_L \equiv  \mbox{\boldmath$r$}\times\left[(\mbox{\boldmath$\mu$}\cdot\nabla)\mbox{\boldmath$B$} - \frac{1}{c}\left( \dot{\mbox{\boldmath$\mu$}} \times \mbox{\boldmath$E$}\right)\right], 
\label{LdotModern_c1}
\end{equation}

where \mbox{\boldmath$B$} here is the magnetic field in the electron rest frame due to the motion of the proton charge around the electron in this frame. (Although the need here is to obtain \(\dot{\mbox{\boldmath$L$}}\) in the laboratory frame, it was noted in Section VB1 that the torque on the orbit calculated in the electron rest frame is equal to the torque in the laboratory frame, provided that \(\gamma\) may be approximated as unity.) It was also shown in Section VB1, for \(\mbox{\boldmath$B$}\) due to the motion of the proton charge in the electron rest frame, that

\begin{equation}
\mbox{\boldmath$r$}\times(\mbox{\boldmath$\mu$}\cdot\nabla)\mbox{\boldmath$B$} = \frac{e}{c r^3} \mbox{\boldmath$r$}\times\left[\mbox{\boldmath$v$} \times \left(3(\mbox{\boldmath$\mu$}\cdot\mbox{\boldmath$n$})
\mbox{\boldmath$n$} - \mbox{\boldmath$\mu$}\right)\right].
\label{TorqueModern_c1}
\end{equation}

(Alternatively, the torque on the orbit could have been calculated in the electron rest frame based on the Biot-Savart force on the proton transiting the intrinsic magnetic field of the electron, obtaining the same result \cite{LushY7}.   The torque must also be obtainable in the laboratory frame.  In this case there is no magnetic field due to the proton, but the electron feels a Stern-Gerlach-like force due to the anisotropy of the proton Coulomb field, on the intrinsic electric dipole moment the electron acquires due to its motion with intrinsic magnetic moment \cite{Fisher71}. The spin-orbit interaction energy can be calculated based on this force \cite{MunozY1} and obtains the expected result. Mu\~noz also notes the necessity of incorporating the hidden momentum in the analysis even in the case of intrinsic magnetic moments of otherwise point-like particles, as opposed to classical magnetic moments due to extended current distributions.)  With \(\mbox{\boldmath$E$}\) the Coulomb field of the proton, Equation (\ref{LdotModern_c1}) becomes

\begin{equation}
\dot{\mbox{\boldmath$L$}} = \frac{e}{c r^3} \mbox{\boldmath$r$}\times\left[\mbox{\boldmath$v$} \times \left(3(\mbox{\boldmath$\mu$}\cdot\mbox{\boldmath$n$})
\mbox{\boldmath$n$} - \mbox{\boldmath$\mu$}\right)- \left( \dot{\mbox{\boldmath$\mu$}} \times \mbox{\boldmath$r$}\right)\right].
\label{TorqueModern_c1}
\end{equation}

Expanding the vector triple products obtains

\begin{widetext}

\begin{equation}
\dot{\mbox{\boldmath$L$}} = \frac{e}{c r^3} \left[ 3(\mbox{\boldmath$\mu$}\cdot\mbox{\boldmath$n$})\left(r\mbox{\boldmath$v$} - \mbox{\boldmath$n$}\left(\mbox{\boldmath$r$} \cdot 
\mbox{\boldmath$v$}\right) \right) -  \left(\mbox{\boldmath$v$}\left(\mbox{\boldmath$r$} \cdot \mbox{\boldmath$\mu$}\right) - \mbox{\boldmath$\mu$}\left(\mbox{\boldmath$r$} \cdot \mbox{\boldmath$v$}\right)\right) - \left(r^2\dot{\mbox{\boldmath$\mu$}} - \left(\dot{\mbox{\boldmath$\mu$}} \cdot \mbox{\boldmath$r$}\right)\mbox{\boldmath$r$}  \right)\right],
\label{TorqueModern_c1}
\end{equation}

and noting that \(\mbox{\boldmath$n$} \cdot 
\mbox{\boldmath$v$} \equiv \dot{r}\),  

\begin{equation}
\dot{\mbox{\boldmath$L$}} = \frac{e}{c r^3} \left[ 3(\mbox{\boldmath$\mu$}\cdot\mbox{\boldmath$n$})\left(r\mbox{\boldmath$v$} - \dot{r}\mbox{\boldmath$r$} \right) -  \left(\mbox{\boldmath$v$}\left(\mbox{\boldmath$r$} \cdot \mbox{\boldmath$\mu$}\right) - r \dot{r}\mbox{\boldmath$\mu$}\right) - \left(r^2\dot{\mbox{\boldmath$\mu$}} - \left(\dot{\mbox{\boldmath$\mu$}} \cdot \mbox{\boldmath$r$}\right)\mbox{\boldmath$r$}  \right) \right].
\label{TorqueModern_c1}
\end{equation}

Further reducing and expressing in terms of the spin, with \(g=2\), arrives at the final form needed as

\begin{equation}
\dot{\mbox{\boldmath$L$}} = -\frac{e^2}{m c^2 r^3} \left[ 2(\mbox{\boldmath$s$}\cdot\mbox{\boldmath$r$})\mbox{\boldmath$v$} - 3(\mbox{\boldmath$s$}\cdot\mbox{\boldmath$r$}) \dot{r}\mbox{\boldmath$n$} + r\dot{r}\mbox{\boldmath$s$} - r^2\dot{\mbox{\boldmath$s$}} + \left(\dot{\mbox{\boldmath$s$}} \cdot \mbox{\boldmath$r$}\right)\mbox{\boldmath$r$} \right].
\label{LdotModernFinal}
\end{equation}

The motion of the spin angular momentum vector assuming \(g=2\) was obtained previously in Equation (\ref{sdotfromThomas6.71}), and can be expressed as

\begin{equation}
\dot{\mbox{\boldmath$s$}}  =  \frac{e^2}{2 m^2 c^2 r^3}\mbox{\boldmath$L$} \times \mbox{\boldmath$s$} =  \frac{e^2}{2 m c^2 r^3}\left( \mbox{\boldmath$r$} \times \mbox{\boldmath$v$}\right) \times \mbox{\boldmath$s$} = \frac{e^2}{2 m c^2 r^3}\left[\left( \mbox{\boldmath$r$} \cdot \mbox{\boldmath$s$}\right)  \mbox{\boldmath$v$} -  \left( \mbox{\boldmath$v$} \cdot \mbox{\boldmath$s$}\right)  \mbox{\boldmath$r$} \right].
\label{s_dot_final}
\end{equation}

Expanding the vector triple product of Equation (\ref{L_hidden_def}) and reducing, the hidden orbital angular momentum may be rewritten as

\begin{equation}
\mbox{\boldmath$L$}_{\text{hidden}}  =  \frac{e}{c r^3}\left[ r^2\mbox{\boldmath$\mu$} - \mbox{\boldmath$r$} \left(\mbox{\boldmath$\mu$} \cdot \mbox{\boldmath$r$}\right) \right].
\label{L_hidden_red}
\end{equation}

The time rate of change of the hidden orbital angular momentum is then obtained as

\begin{equation}
\dot{\mbox{\boldmath$L$}}_{\text{hidden}}  =  \frac{e}{c r^3}\left[ r^2\dot{\mbox{\boldmath$\mu$}} - \mbox{\boldmath$v$} \left(\mbox{\boldmath$\mu$} \cdot \mbox{\boldmath$r$}\right) - \mbox{\boldmath$r$} \left(\left(\dot{\mbox{\boldmath$\mu$}} \cdot \mbox{\boldmath$r$}\right) + \left(\mbox{\boldmath$\mu$} \cdot \mbox{\boldmath$v$}\right)\right)  - r \dot{r} \mbox{\boldmath$\mu$} + 3\dot{r}\mbox{\boldmath$n$} \left(\mbox{\boldmath$\mu$} \cdot \mbox{\boldmath$r$}\right)  \right].
\label{L_hidden}
\end{equation}

Rewriting the electron intrinsic magnetic moment in terms of the spin, assuming \(g=2\), and rearranging terms for convenience obtains the final form for the rate of change of the hidden orbital angular momentum as

\begin{equation}
\dot{\mbox{\boldmath$L$}}_{\text{hidden}}  =  \frac{e^2}{m c^2 r^3}\left[  \mbox{\boldmath$v$} \left(\mbox{\boldmath$s$} \cdot \mbox{\boldmath$r$}\right) + \mbox{\boldmath$r$}\left(\mbox{\boldmath$s$} \cdot \mbox{\boldmath$v$}\right)  + r\dot{r} \mbox{\boldmath$s$} - 3\dot{r}\mbox{\boldmath$n$} \left(\mbox{\boldmath$s$} \cdot \mbox{\boldmath$r$}\right)  - r^2\dot{\mbox{\boldmath$s$}} + \left(\dot{\mbox{\boldmath$s$}} \cdot \mbox{\boldmath$r$}\right)\mbox{\boldmath$r$}\right].
\label{L_hidden_dot_final}
\end{equation}

Summing Equations (\ref{LdotModernFinal}), (\ref{s_dot_final}), and (\ref{L_hidden_dot_final}) in accordance with Equation (\ref{MotionTotMechAngMomWithHidden}) then obtains the rate of change of the total angular momentum as

\begin{equation}
\dot{\mbox{\boldmath$J$}} = -\frac{e^2}{2m c^2 r^3} \left[ \left( \mbox{\boldmath$r$} \cdot \mbox{\boldmath$s$}\right)  \mbox{\boldmath$v$} -  \left( \mbox{\boldmath$v$} \cdot \mbox{\boldmath$s$}\right)  \mbox{\boldmath$r$}  \right]= \frac{e^2}{2m c^2 r^3}  \mbox{\boldmath$s$} \times \left( \mbox{\boldmath$r$} \times \mbox{\boldmath$v$}\right)= \frac{e^2}{2m^2 c^2 r^3}  \mbox{\boldmath$s$} \times \mbox{\boldmath$L$}.
\label{MotionTotAngMomFinal}
\end{equation}

\end{widetext}

The essential motion of the total angular momentum described by Equation (\ref{MotionTotAngMomFinal}), even absent an external field, is exactly attributable to the Thomas precession of the electron spin angular momentum vector.  Otherwise, if the Thomas precession is negligible, the total angular momentum is a constant of the motion.  This is in contrast to the case of the total angular momentum omitting the hidden angular momentum, as considered by Thomas, which is only generally a constant of the motion in the orbit-averaged, or secular, sense, even absent Thomas precession.

Kiessling \cite{Kiessling1998} has shown that spin is a necessity in classical electron theory, in order to obtain angular momentum conservation, ``independently of the Thomas precession''.  In this light it should probably not be taken for granted that angular momentum will be conserved in a system consisting of a classical current and a particle possessing intrinsic spin and nonclassical gyromagnetic ratio, even absent Thomas precession.  However, the coexistence of classical and nonclassical gyromagnetic ratios in the dynamics does not present a problem to angular momentum conservation according to the present analysis.  It is only the Thomas precession that causes the violation.

\section{Discussion of the prospect for a consistent simultaneous description of the spin-orbit coupling and the anomalous Zeeman effect in classical spin theory}

Apparently, the main objective of Thomas's 1927 paper, that is, to consistently explain the magnitude of the spin-orbit coupling simultaneously with that of the anomalous Zeeman effect, is not met by either Thomas's or the present analysis.  The problem of Thomas's analysis, unrecognized at the time and until decades later, is that linear momentum is nonconserved when the hidden momentum is omitted.  As the present analysis shows, on the other hand, when the hidden momentum is incorporated in its equation of motion, the  angular momentum is no longer even secularly conserved. 

The problem that results in either nonconservation case is one of building a consistent description of the spin-orbit coupling. Given linear or angular momentum nonconservation, it should not be surprising that in neither of these cases is there a single unique law derivable for the spin-orbit coupling.  Different coupling values will be obtained depending on the focus of the particular calculation used.  For example, one may calculate the binding force on either particle due to the other, and the influence of the relative orientation of the spin or orbit, as applicable, on its magnitude.  The binding force here is the Coulomb attraction modified slightly by the electric field that results from the orbital motion of the magnetic electron. If the binding force is not equal and opposite on each particle, there is not a single binding energy value that results.  Rather, there are two.  This is the result for Thomas's analysis that omitted the hidden momentum, as would be expected according to the modern textbooks.

Alternatively, the energy involved in spatially inverting the plane of the orbit, thus reversing the direction of the orbital angular momentum vector, or of inverting the spin, may be calculated. According to the present analysis, the spin angular momentum vector is precessing such that the magnitude of its rate of change is half the magnitude of the orbital angular momentum rate of change.  Since the rate of change of angular momentum equates to the applied torque, the amount of work done in inverting the spin angular momentum must be half that done in inverting the orbital plane.  Yet, this cannot be because (assuming the absence of applied field) the binding energy before and after the inversion cannot depend on the orientation of the system. In the absence of external field these must not differ, yet they will if the equation of torque and rate of change of angular momentum is assumed to hold even in the presence of Thomas precession.

It is tempting to suppose that since the Thomas precession is a kinematic effect, in its presence the standard relation between a torque and the change in angular momentum no longer applies.  This at least obtains a consistent value for  the spin orbit coupling, since the Thomas precession does not otherwise affect the dynamics.  However, if this is the case, then the Thomas precession  fails to account for the spin-orbit coupling anomaly of one-half, as it can have no effect on the coupling magnitude.  If it is nonetheless accepted that the anomalous Zeeman effect implies the electron g-factor is two, then the spin-orbit coupling will be twice its measured value.  This is arriving essentially back at the situation for classical spin theory prior to Thomas's paper of 1927.   

As the quote in Section I herein shows, Thomas recognized that secular angular momentum conservation was a necessary condition for a consistent simultaneous description of spin-orbit coupling and the anomalous Zeeman effect.  Therefore the present results are not entirely inconsistent with Thomas's.

\section{Character of the motion of the total angular momentum}

Having established that the total angular momentum cannot be stationary for nonaligned spin and orbital angular momenta, it is of interest to determine the nature of the motion.  It is assumed again for simplicity that the orbit is circular so that \(\omega_L\) and \(\omega_s\) are constants of the motion.

\subsection{Constancy of total angular momentum magnitude}

The total angular momentum magnitude is constant if \(d(J^2)/dt=0 \), where 

\begin{equation}
J^2 = \mbox{\boldmath$J$} \cdot \mbox{\boldmath$J$} = L^2 + s^2 + 2 \mbox{\boldmath$L$} \cdot \mbox{\boldmath$s$}.
\nonumber
\end{equation}

The orbital and spin angular momenta magnitudes are constant, so \(d(J^2)/dt=0 \) is equivalent to

\begin{equation}
\dot{\mbox{\boldmath$L$}} \cdot \mbox{\boldmath$s$} = -\mbox{\boldmath$L$} \cdot \dot{\mbox{\boldmath$s$}},
\nonumber
\end{equation}

or

\begin{equation}
(\mbox{\boldmath$L$} \times \mbox{\boldmath$\omega$}_L) \cdot \mbox{\boldmath$s$} = -\mbox{\boldmath$L$} \cdot (\mbox{\boldmath$s$} \times \mbox{\boldmath$\omega$}_s),
\nonumber
\end{equation}

which, by the properties of the scalar triple product can be rewritten as

\begin{equation}
\mbox{\boldmath$L$} \times \mbox{\boldmath$\omega$}_L \cdot \mbox{\boldmath$s$} = \mbox{\boldmath$L$} \times \mbox{\boldmath$\omega$}_s \cdot \mbox{\boldmath$s$} .
\nonumber
\end{equation}

Taking account that the spin and orbit precess around each other obtains 

\begin{equation}
\mbox{\boldmath$L$} \times \omega_L \hat{\mbox{\boldmath$s$}} \cdot \mbox{\boldmath$s$} = \mbox{\boldmath$L$} \times \omega_s \hat{\mbox{\boldmath$L$}} \cdot \mbox{\boldmath$s$}. 
\nonumber
\end{equation}

Both sides of this equation are identically zero by the properties of the scalar triple product, so the equality is satisfied for all \(\mbox{\boldmath$L$} \) and \( \mbox{\boldmath$s$}\).  The magnitude of the total angular momentum is thus a constant of the motion for all relative magnitudes and orientations of \(\mbox{\boldmath$L$} \) and \( \mbox{\boldmath$s$}\), and for all electron-proton separations.

\subsection{Angular velocity of the total angular momentum}

It has been established that although the vector total angular momentum cannot be constant in the current model, the total angular momentum magnitude, \(J\), is constant at all orbital radii.  Given that \(J\) is constant, the time derivative of the total angular momentum can be written generally as

\begin{equation}
\mbox{\boldmath$L$} \times \omega_L \hat{\mbox{\boldmath$s$}} +
\mbox{\boldmath$s$} \times \omega_s \hat{\mbox{\boldmath$L$}} = \mbox{\boldmath$J$} \times \mbox{\boldmath$\omega$}_J,
\nonumber
\end{equation}

where \(\omega_L\) and \(\omega_s\) are constant (and given by Eqs. (\ref{Omega_s}) and (\ref{LAngVel})  below) for a fixed orbit radius, and where \( \mbox{\boldmath$\omega$}_J \) is a strictly angular velocity.  Equivalently, since \(\mbox{\boldmath$J$} \equiv \mbox{\boldmath$L$} + \mbox{\boldmath$s$}\),

\begin{equation}
\mbox{\boldmath$J$} \times \omega_L \hat{\mbox{\boldmath$s$}} +
\mbox{\boldmath$J$} \times \omega_s \hat{\mbox{\boldmath$L$}} = \mbox{\boldmath$J$} \times \mbox{\boldmath$\omega$}_J,
\nonumber
\end{equation}

from which

\begin{equation}
\mbox{\boldmath$\omega$}_J = \omega_L \hat{\mbox{\boldmath$s$}} +
\omega_s \hat{\mbox{\boldmath$L$}}, 
\label{omega_J}
\end{equation}

and

\begin{equation}
\dot{\mbox{\boldmath$\omega$}}_J = \omega_L \dot{\hat{\mbox{\boldmath$s$}}} +
\omega_s \dot{\hat{\mbox{\boldmath$L$}}} = \frac{\omega_L}{s} \dot{\mbox{\boldmath$s$}} +
\frac{\omega_s}{L} \dot{\mbox{\boldmath$L$}} ,
\nonumber
\end{equation}

or

\begin{equation}
\dot{\mbox{\boldmath$\omega$}}_J = \frac{\omega_L}{s} (\mbox{\boldmath$s$} \times \omega_s \hat{\mbox{\boldmath$L$}}) +
\frac{\omega_s}{L} (\mbox{\boldmath$L$} \times \omega_L \hat{\mbox{\boldmath$s$}}) ,
\nonumber
\end{equation}

or

\begin{equation}
\dot{\mbox{\boldmath$\omega$}}_J = (\omega_L \omega_s - \omega_s\omega_L ) \hat{\mbox{\boldmath$L$}} \times  \hat{\mbox{\boldmath$s$}} \equiv 0.
\nonumber
\end{equation}

Thus, the total angular momentum precesses at a constant rate around a fixed axis, as illustrated on Figure 2.  This type of motion, occurring even absent an external magnetic field \cite{EisbergResnick} (except that the precession is said to be ``random''), is thought to be essentially quantum mechanical.

\begin{figure}
	\centering
		\includegraphics[width=0.4\textwidth]{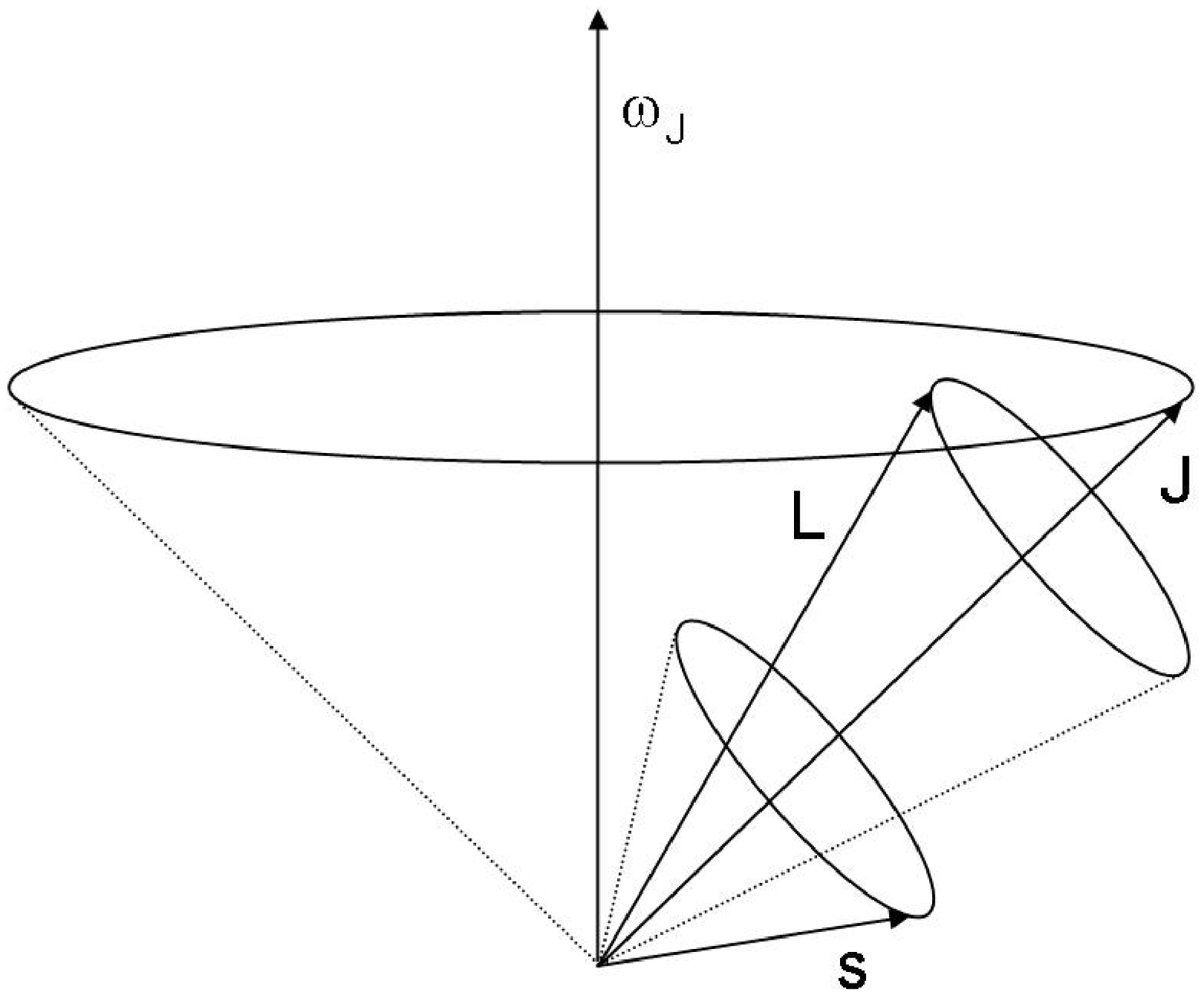}
	\caption{Present Thomas precession, \(\mbox{\boldmath$J$}\) must precess if  \(\mbox{\boldmath$L$}\) and \(\mbox{\boldmath$s$}\) are not aligned. (After \protect\cite{EisbergResnick}).}
	\label{fig:J_precesses}
\end{figure}

\subsection{Magnetic nonradiativity when \(g=2\)}

The intensity of radiation due to a magnetic dipole of moment, \(\mbox{\boldmath$m$}\), is \cite{LandauLifshitzClassToF}

\begin{equation}
I = \frac{2}{3c^3}|\ddot{\mbox{\boldmath$m$}}|^2.
\label{MagDipoleRadIntensity}
\end{equation}

The condition of being nonradiative due to magnetic dipole moment motion is thus that \(\ddot{\mbox{\boldmath$m$}} = 0\).  

The total of the intrinsic and orbital magnetic moments is

\begin{equation}
\mbox{\boldmath$m$} = \mbox{\boldmath$\mu$} +
\frac{e}{2 m c }\mbox{\boldmath$L$},
\label{MagMoment}
\end{equation}

or, in terms of the intrinsic spin,

\begin{equation}
\mbox{\boldmath$m$} = \frac{g e}{2 m c } \mbox{\boldmath$s$} +
\frac{e}{2 m c}\mbox{\boldmath$L$}.
\label{MagMoment}
\end{equation}

Then

\begin{equation}
\dot{\mbox{\boldmath$m$}} = \frac{g e}{2 m c } \dot{\mbox{\boldmath$s$}} +
\frac{e}{2 m c}\dot{\mbox{\boldmath$L$}}.
\label{MagMoment}
\end{equation}

From Eq. (\ref{SpinEOMAnyg}) with \(\dot{ \mbox{\boldmath$s$}} \equiv \mbox{\boldmath$\omega$}_s \times \mbox{\boldmath$s$}\), it is obtained that

\begin{equation}
\mbox{\boldmath$\omega$}_s = \left(\frac{g}{2} - \frac{1}{2} \right)\left(\frac{e^2 L}{m^2c^2 r^3}\right) \hat{\mbox{\boldmath$L$}} 
\label{Omega_s}
\end{equation}

is the angular velocity of the spin axis around the orbit normal. The angular velocity of the orbit normal around the spin axis, based similarly on Eq. (\ref{ModernLdot}), is found to be

\begin{equation}
\mbox{\boldmath$\omega$}_{L} = \frac{g e^2 s}{2m^2c^2 r^3}  \hat{\mbox{\boldmath$s$}}.
\label{LAngVel}
\end{equation}

Thus,

\begin{equation}
\dot{\mbox{\boldmath$m$}} = \frac{g e}{2 m c } \mbox{\boldmath$s$} \times \frac{\omega_s}{L} \mbox{\boldmath$L$} +
\frac{e}{2 m c}\mbox{\boldmath$L$} \times \frac{\omega_L}{s} \mbox{\boldmath$s$}.
\label{MagMoment}
\end{equation}

Taking a second derivative with respect to time obtains

\begin{equation}
\ddot{\mbox{\boldmath$m$}} = \left(\frac{g e}{2 m c }\frac{\omega_s}{L}- \frac{e}{2 m c}\frac{\omega_L}{s} \right)\left[ \dot{\mbox{\boldmath$s$}} \times  \mbox{\boldmath$L$} + \mbox{\boldmath$s$} \times \dot{\mbox{\boldmath$L$}} \right]. 
\label{MagMomentDDVanish}
\end{equation}

Now, \( \ddot{\mbox{\boldmath$m$}}\) will vanish if either the quantity in parentheses or that in brackets vanishes.  Vanishing the former requires that

\begin{equation}
gs\omega_s = L\omega_L, 
\label{MagMomentDDVCond}
\end{equation}

or, using again Eqs. (\ref{Omega_s}) and (\ref{LAngVel}) for \(\omega_L\) and \(\omega_s\), 

\begin{equation}
gs(g-1)L = Lgs, 
\label{MagMoment}
\end{equation}

or

\begin{equation}
g = 2. 
\label{MagMoment}
\end{equation}

So, \(g=2\) obtains a general nonradiative condition, independent of the relative orientation of the spin and orbit.  It must be noted, however, that this cannot be taken as an exact result based on the present analysis where the relativistic gamma factor is approximated as unity and, more importantly, the finite and specific mass of the proton is not taken into account.

It is not difficult to show that the quantity in square brackets in Eq. (\ref{MagMomentDDVanish}) does not generally vanish.

Malykin \cite{MalykinY6} states, ``The analysis in [Bagrov, {\em et al.} \cite{BagrovY2}] suggests that the
TP cannot be regarded as the source of relativistic radiation
power.''  The present analysis, in disagreement, suggests that present Thomas precession, magnetic dipole radiation will occur unless the gyromagnetic ratio has a specific nonclassical value.

\section{Summary of findings}

The following observations are supported by the analyses provided:

1) L. H. Thomas's paper of 1927 did not successfully demonstrate that his ``relativity precession'' could account for the anomalous factor of one-half in atomic spin-orbit coupling magnitude, while simultaneously explaining the anomalous Zeeman effect.  This conclusion must be drawn because Thomas's dynamical equations omitted the effect of the hidden momentum, which has been recognized in recent decades to be necessary in order to achieve linear momentum conservation in the electrodynamic interaction of point charges and magnetic dipoles.  When linear momentum is nonconserved, Newton's law of action and reaction is violated, and the binding force and binding energy cannot be consistently calculated in the two-body system where the bodies are oppositely charged and one has an intrinsic magnetic moment.

2) Angular momentum cannot be conserved in a system consisting of two bound oppositely-charged particles where one posseses an intrinsic magnetic moment, when the Thomas precession is not insignificant, according to the modern electrodynamical picture that incorporates the effect of the hidden momentum.  The Thomas precession thus cannot account successfully for the spin-orbit coupling anomaly either using the equation of translational motion of a magnetic dipole according to the modern textbooks.  

3) Despite angular momentum nonconservation, the total magnetic moment of the quasiclassical hydrogenic atom can be generally stationary if the magnetic dipole has a specific nonclassical gyromagnetic ratio.  Then, despite the motion of the total angular momentum, the system will not emit magnetic dipole radiation.  The gyromagnetic ratio required for magnetic dipole nonradiativity corresponds to a g-factor of approximately 2.

\section{Acknowledgement}

I thank Paul de Haas for his  \href{http://home.tiscali.nl/physis/HistoricPaper/}{Physis Project} website, where L. H. Thomas's paper of 1927 may be downloaded.

\bibliographystyle{plain}

%\bibliography{hydrogen}

\end{document}